\documentclass{aa}  

\usepackage{graphicx}
\usepackage{multirow}%
\usepackage{amsmath,amssymb,amsfonts}%
\usepackage{mathrsfs}%
\usepackage[title]{appendix}%
\usepackage{xcolor}%
\usepackage{textcomp}%
\usepackage{manyfoot}%
\usepackage{booktabs}%
\usepackage{algorithm}%
\usepackage{algorithmicx}%
\usepackage{algpseudocode}%
\usepackage{listings}%
\usepackage{ulem}
\usepackage{txfonts}

\newcommand{\xmm}{\textit{XMM-Newton }}

\begin{document}

	\title{Probing the star-forming environment  of  GRB across  cosmic ages with high resolution X-ray spectroscopy}
	\subtitle{From \textit{{XMM-Newton}}/RGS and \textit{XRISM}/Resolve to \textit{NewAthena}/X-IFU}
	
	
	\author{Luigi Piro
		\inst{1}
		\and
		Aishwarya Linesh Thakur \inst{1}
		\and
		Alfredo Luminari
		\inst{2}
		\and
		Fabrizio Nicastro
		\inst{2}
		\and
		Yair Krongold
		\inst{3}
	}
	
	\institute{
		INAF -- Istituto di Astrofisica e Planetologia Spaziali, via Fosso del Cavaliere 100, I-00133 Rome, Italy\\
		\email{luigi.piro@inaf.it}
		\and
		INAF - Osservatorio Astronomico di Roma, Via Frascati 33, 00078 Monteporzio, Italy
		\and
		Instituto de Astronomía, Universidad Nacional Autónoma de México, Circuito Exterior, Ciudad Universitaria, Ciudad de México 04510, México}
	
	\date{Received; accepted}
	\abstract
	{Optical and infrared absorption spectroscopy of GRB afterglows provides valuable insight into host-galaxy properties across cosmic time. However, the immediate circumstellar environment---the star-forming region embedding the progenitor on scales $\lesssim 100~\mathrm{pc}$ is heavily ionized by the GRB prompt and afterglow emission, making X-ray spectroscopy the sole viable probe of its physical and chemical conditions.}
	{We characterize the GRB-embedding medium as a function of progenitor type and cosmic epoch via high-resolution X-ray spectroscopy. We self-consistently compute narrow X-ray absorption lines from various elements and ionization stages in star-forming regions hosting long GRBs. We extend our study to high-redshift GRBs to evaluate how effectively X-ray absorption signatures can distinguish between pristine, Pop~III-, and Pop~II-enriched environments.}
	{Using the time-dependent photoionization code \texttt{TEPID}, we calculate the ionization structure of the absorber along with the energies and strengths of the resulting absorption lines. We analyze a flux-selected sample of GRBs observed by XMM-Newton/PN that show strong evidence for time-dependent and layered photoionization, searching for predicted narrow features in the corresponding RGS spectra. Finally, we assess the performance of \textit{XRISM}/Resolve and \textit{NewAthena}/X-IFU in detecting these features and constraining key absorber properties, such as redshift, ionization state, and elemental abundances.}
	{In the single archival case with sufficient RGS signal-to-noise ratio (GRB~090618 at $z=0.54$), we find $\approx 3\sigma$ combined evidence for narrow absorption lines, most prominently from highly ionized Fe~L. For bright nearby GRBs, \textit{XRISM}/Resolve can detect Fe~K absorption lines even in its current gate-valve-closed configuration ($\sim 3$ expected events/year). Looking ahead, \textit{NewAthena}/X-IFU will enable systematic population studies of GRB environments, delivering redshifts with precision comparable to optical/IR spectroscopy, unaffected by dust obscuration and Ly-$\alpha$ forest absorption at high redshift.  Crucially, it will provide precise diagnostic measurements of density, size, metallicity, and abundance patterns, ultimately revealing whether high-$z$ GRBs originate in pristine, Pop~III-, or Pop~II-enriched media.}
	{}
	
	
	\maketitle

	\section{Introduction}
	Long duration gamma-ray bursts (hereafter GRBs) are primarily associated with the death of massive stars. This evidence comes from a combination of various facts. They harbor primarily in dwarf irregular or sub-luminous star-forming galaxies (\cite{Blanchard2016, Lyman2017}), being located within the brightest UV (star-forming) regions of the galaxy (\cite{Svensson2010, Fruchter2006}). Furthermore, but for a few recent exceptions, they were found since the early discoveries of afterglows to be associated with supernovae type Ic (\cite{Galama1998, Stanek2003}). These observational facts clearly point to a close relationship between star formation and GRBs, suggesting that the ambient medium in which GRBs explode is denser than the interstellar medium and typical of star forming regions.
	
	Spectroscopy in the optical band has enabled extensive studies of the interstellar
	medium (ISM) of the GRB host galaxy in absorption \cite[e.g.][]{2006ApJ...648...95P} and the properties of the host itself in emission (\cite{2015A&A...581A.125K}). Absorption in the observed optical-UV spectra arises from low- to moderate-ionisation ultraviolet metals (eg. Mg I-II, CIV and NV) in the host ISM \citep{2008A&A...491..189F, 2018MNRAS.479.3456H, Saccardi2026}. Analysis of time-varying, excited fine-structure lines of these metals places this material at sub-kpc to kpc scales from the GRB site \citep{Vreeswijk2007, DElia2010}. However, the 1-100 pc region embedding the burst has been so far unexplored, as extremely high ionisation by the GRB precludes optical absorption on such distances.
	
	An increasingly crucial role in diagnosing the close environment of GRB has emerged with the observation of X-ray absorption. Evidence of X-ray absorption in excess to that of our Galaxy was found since the very first X-ray afterglows were observed  with \textit{BeppoSAX} \citep{Owens+98}. A systematic analysis of the \textit{BeppoSAX} afterglow sample  \citep{Stratta+2004,Depasquale+2006} indicated intrinsic (i.e. rest-frame) column density in the range $N_H\approx (0.1\sim10)\times 10^{22}\mathrm{cm}^{-2} $, clustering predominantly around a few $\times10^{22}\ \mathrm{cm}^{-2}$. X-ray absorption is produced predominantly by metals, but is is customary to quote it in hydrogen column density for solar-like abundances. A comparable range was obtained  in a sample of afterglows observed by \textit{XMM-Newton} and \textit{Chandra} \citep{Gendre+06}.  Systematic observations by \textit{Swift} increased the sample, confirming that the afterglows are heavily absorbed around a mean value $N_H=5\times 10^{22}\mathrm{cm}^{-2} $ \citep{Campana+2012}. Furthermore, the distribution of $N_H$ is consistent with the observed properties of Giant Molecular Clouds in our Galaxy (GMCs) \citep{Campana+10}.
	
	A comparison between optical and X-ray column densities provides a further important hint as to the location of such absorbing medium. X-ray absorbing columns are always larger than optical columns  by a factor of $\gtrsim 10$ \citep{Schady+10}. This can be interpreted as a manifestation of GRB photoionization, locating the X-ray absorbing material close to the GRB. The luminous GRB emission is expected to have a strong ionising effect on the gas situated within 100 parsec of the GRB site. In this region, the gas is ionised to a degree that it becomes transparent to optical-UV photons. On the other hand soft X-ray absorption remains effective even in such large ionizing environment, because  metals  (and, to some extent, He, \cite{Watson+10}) are not expected to be completely ionized even as close to the GRB. A systematic comparison of the optical and X-ray column densities have found an indication for highly photoionised material within a few parsec of the GRB site \citep{Watson+10, KP13}. Overall, all acquired evidences strongly suggest that the X-ray absorption arises in the photoionised medium of the star-forming region hosting the GRB progenitor.
	
	However, contrary to those expectations, photoionisation-equilibrium modelling of the X-ray afterglow spectra of GRBs have generally resulted in poor fits, either
	returning results consistent with a neutral medium  or leaving the ionisation parameter totally unconstrained \citep{Piro+02}. The issue arises because photoionised absorber models generally assume photoionisation equilibrium in the medium and slab geometry. This assumption is not appropriate to properly model both the strong time-dependent ionizing flux from the GRB and the ionization stratification in distance  in the medium close to the GRB. These conditions are fully developed in our TEPID model. \citep{Luminari2023}. In \cite{Thakur2026} (hereafter Paper I) we have first applied the model to the \xmm EPIC-PN data of a sample of bright GRB afterglows with known redshift.  We find that the model provides a good fit to the spectra, strongly preferred to a neutral absorber in several cases. The unique time- and spatially-dependent ionisation structure of the gas imprinted on the X-ray absorbing {\it continuum}  determines independently the density $(log(n)\approx 2-4)$ and size $r\approx 5-100 pc$,  allowing us to locate the absorbing medium of this representative sample of long GRBs in the region of the density-size diagram populated by star-forming regions.
	
	A rich ensemble of narrow absorption lines are expected by the model. However, although the XMM-PN is sensitive to the distinct shape of the absorption continuum imprinted by the structured, non-equilibrium, ionized absorbers,  its energy resolution is not sufficient to detect the  lines. This requires instrument with energy resolution at the level of few eV. Past studies have attempted {\it blind searches } for emission and absorption features, \citep[see][]{Campana2011, Campana2016, Campana2024} in \textit{XMM-Newton} RGS spectra of high-fluence GRBs. These studies find marginal evidence for emission features, albeit statistically not significant, and no significant absorption features. 
	
	In this paper we thus expand on the results of Paper I and explore the capabilities of high-resolution X-ray spectroscopy instruments to carry out {\it a targeted search} of the rich ensemble of   all the broad and narrow line features {\it predicted} by the TEPID model, and leading to precise measurements of the density, size, and metal abundances of the star-forming medium embedding the GRB.
	In Sect.\ref{sec:TEPID} we first summarize the TEPID model and introduce the updates in the line database that have been implemented for this paper.
	Following  the model results as derived in Paper I for the \xmm sample, we discuss in Sect.\ref{sec:predicting} the associated rich spectrum of narrow lines and present a search of these features in the \xmm RGS spectra obtained during the same observations in Sect.\ref{RGS}.  

	We then estimate the prospects of observing bright GRB afterglows and detect  narrow absorption lines in close and bright GRB afterglows with the Resolve instrument onboard \textit{XRISM} (\cite{2021SPIE11444E..22T}, Section \ref{XRISM}).
	
	With its unique sensitivity to weak and narrow lines,  resulting from the energy resolution of the non-dispersive microcalorimenter at the core  of the XIFU instrument \citep{2023ExA....55..373B, 2025ExA....59...18P} combined with the large collecting area of the telescope, \textit{NewAthena} is expected to be a game-changer in this field \citep{Piro+22}. In Sect. \ref{NewAth} we gauge the expected improvement by extending the analysis from  the XMM sample of GRB afterglows to higher redshift and fainter GRBs. We showcase \textit{NewAthena}'s capability to routinely capture high-resolution spectra for numerous GRB afterglows. This allows us to probe the unique signatures of the high-redshift star-forming medium across cosmic time, tracing its evolution from Population II solar-like gas to Population III-enriched environments, and ultimately to the pristine gas from which the first Population III stars originated. In Sect. \ref{Conclusions} we draw our conclusions.
	
	\section{Ionized absorbers in the close environment of GRBs}
	\label{sec:ionizedabsorbers}
	\subsection{ Time Evolving Photoionization in GRBs}\label{sec:TEPID}
	The circumburst environment of a GRB is far from ionisation equilibrium. The extremely large rate of ionizing photons produced by a GRB quickly ionizes the surrounding medium at a rate much higher than it can recombine, for any reasonable density: while much of the fluence is accumulated in the first $10^2-10^3$ seconds, typical recombination timescale for a gas with $n\sim 10^4 cm^{-3}$ and T=100 Kelvin (as expected for star-forming regions around GRBs; see below) is of the order of years and increases linearly with lower densities. Therefore, to meaningfully characterize the surrounding gas and, thus properly model the absorption spectrum and fit it to observed data, a proper physical model is needed. The Time-Evolving PhotoIonisation Device (TEPID; \citealp{Luminari2023}) does exactly this: it models the physics of gas illuminated by a time-variable luminosity source and can handle quite different scenarios, from variable AGNs (e.g. \citealp{Serafinelli2025, Luminari2026}) to transients like long GRBs (e.g. Paper I).
	
	Since the first version presented in \cite{Luminari2023}, TEPID has been continuously updated. In view of the higher energy resolution of current and future microcalorimeter missions such as \textit{XRISM} and \textit{NewAthena}, it now includes more elements and significant updates to the spectral fitting tool. We now include Fl, Na, Al, P, Cl, Ar, K, Ca, Ti, Mn, Ni, Zn (i.e. all those more abundant than $10^{-7}$); photoionisation cross section are taken from Cloudy \citep{2025RMxAA..61c.120G} and Auger probabilities from \cite{1993A&AS...97..443K}.
	
	Finally, our spectral interface between TEPID and a fitting package (PHASE; \citealp{2003ApJ...597..832K}) now includes analytically-computed Voigt profiles \citep{2006MNRAS.369.2025T} significantly increasing the accuracy of the computed line profiles. The atomic line database has been expanded for resonant transition from heavy elements and several inner-shell resonant transitions of lighter elements with the XSTAR database \citep{2021Atoms...9...12M}. Of these transitions, we only consider those starting from the ground-state L-S multiplets, assuming that they all contribute equally.

	\begin{figure}
		\centering
		\includegraphics[width=\linewidth]{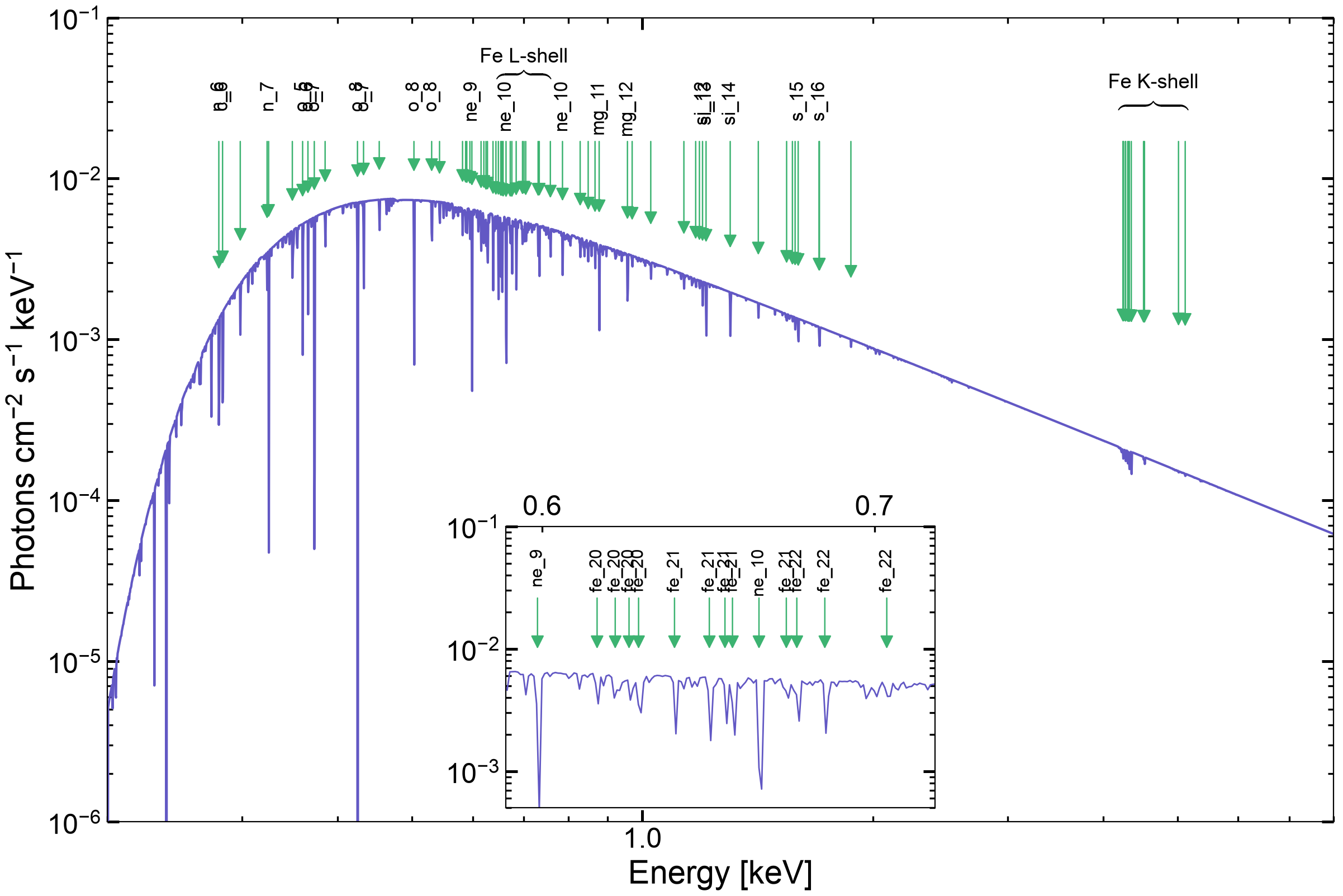}
		\caption{TEPID model spectrum for the best-fit parameters of GRB 090618. To emphasize the richness of the metal lines present in the spectrum, we have removed the Galactic column (5.95$\times 10^{20} \mathrm{cm}^{-2}$), with each individual transition having an observed equivalent width larger than 0.5 eV marked in green. The inset panel zooms-in on the 0.6 - 0.7 keV range, where we can see multiple Fe-L shell transitions.}
		\label{fig:GRB090618_photon}
	\end{figure}

	\begin{figure}
		\centering
		\includegraphics[width=0.5\textwidth]{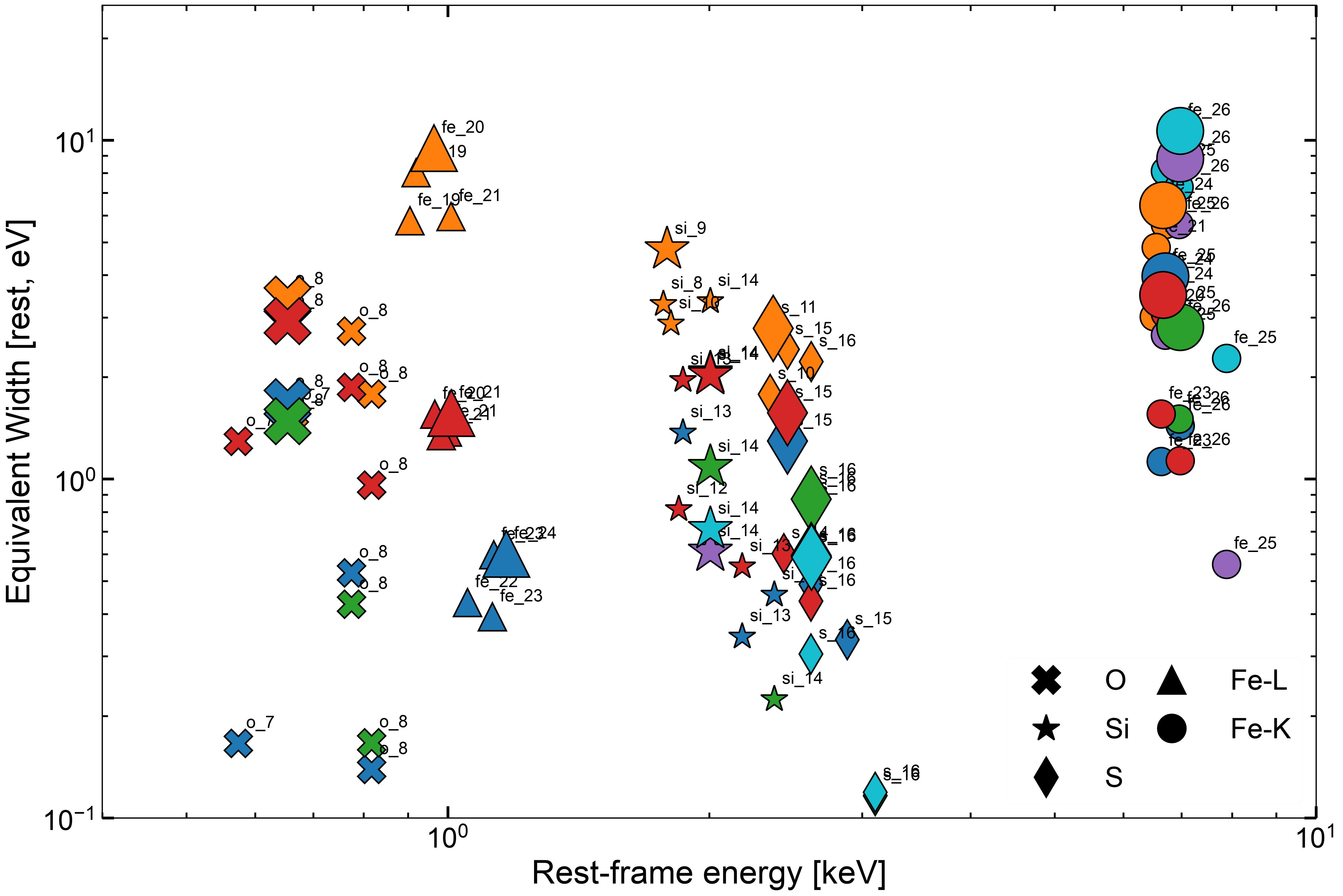}
		\caption{Rest-frame energy versus rest-frame equivalent width for the four strongest absorption lines estimated using the per-GRB best-fit TEPID parameters of the Paper I sample: GRB 060729 (dark blue), GRB 061121 (orange), GRB 080411 (green), GRB 090618 (red), GRB 120711A (magenta), GRB 221009A (cyan). Different symbols plot the absorption lines for silicon (stars), sulfur (diamonds), iron L-shell (triangles) and iron K-shell (circles)}
		\label{fig:EWrest}
	\end{figure}

	\subsection{Predicting narrow absorption lines}
	\label{sec:predicting}
	In the subsample of GRB afterglows observed by XMM-Newton and containing more than $10^5$ total counts in their PN spectra, the TEPID model provides an excellent fit to the PN data, and is  preferred to a cold neutral intrinsic absorber (Paper I). In fig.\ref{fig:GRB090618_photon} we show the TEPID model that best-fits the PN spectrum of GRB~090618. It is rich of narrow absorption lines produced by metals ionized by the GRB and spans a wide ionization range, from He-like carbon all the way up to H-like Fe. The XMM-PN is  sensitive to the distinct shape of the absorption continuum imprinted by the structured, non-equilibrium, ionized absorbers but its energy resolution is not sufficient to detect the single lines.  Given the relative weakness of those  lines, spanning an equivalent width interval of  (EW) $\sim 1-10~$eV, a detection requires instrument with  energy resolution at the level of few eV. 
	
	As noted in Paper I, in a simple single-zone photoionization code  just one or at most two ions of a same element are responsible for most of the bound-bound opacity. On the contrary, a distinct feature of the spectrum of the GRB ionized absorber, when properly modelled by TEPID, is the rather large spread of ionization stages, primarily driven by the dilution of the ionization flux  with distance,  that are effectively producing absorption.   This is clear from
	Fig.\ref{fig:GRB090618_photon}, where we plot as an example the model spectrum of GRB090618. Most prominent absorption lines with EW of several eV from a variety of ions from O, N, Ne, Mg, Si, S , Fe are imprinted on the spectrum. In Fig.\ref{fig:EWrest} we plot, for each GRB of the Paper I sample the rest-frame EW of the 4 most prominent lines, as predicted by our best-fit TEPID models, for a selection of elements with lines above 0.9 keV rest frame, so to cover a wide range of redshifts. 
	
	To emphasize the spread in different bursts, we have identified with the largest symbol the strongest lines for each element. 
	For Fe K lines, the rest-frame  EW ranges from 4 to 10 eV, for S from 0.7 to 3 eV, for Si from 0.6 to 5 eV, while for Fe-L shell only 3 bursts present a line with EW above 0.1 eV. Thus, the weakest lines are typically associated to lower-Z elements or shells with lower ionization potential, indicating that the overall ionization of the absorber is one driving parameter. 
	
	We thus expect relatively strong lines for all GRBs of the sample, namely GRB060729, GRB061121, GRB080411, GRB090618, GRB120711A and GRB 221009A (the so-called BOAT) \footnote{Note that GRB190114C is heavily absorbed below 2 keV by its host galaxy, so it is not included in the present analysis.}.
	
	In the  model adopted to fit the XMM data we assumed solar metallicity.
	This assumption does not affect significantly  the expected strength  of absorption features by metals, that is determined by their column densities. Current X-ray experiments are primarily sensitive to absorption by metals. Expressing their column density in hydrogen equivalent for a solar abundance is adopted for convention.
	As noted in Paper I, some effect of absorption by H and He can be significant up to rest frame energies of $\lesssim 1~$keV (see also \cite{Watson+10}), thus potentially be detectable up to  $z\lesssim 1.5$ for current experiments.  On the other hand,  when   the metallicity is allowed to vary in a range comprised between the host galaxy metallicity (typically $\gtrsim 0.2~Z\odot$  and solar), the column density of metals do not change substantially (Paper I). 
	This happens because X-ray absorption from a region surrounding a GRB explosion results from a stratified ionisation structure that sets the integrated relative contribution of H, He and metal columns. This is to some extent degenerate with  model parameters (density, size), meaning that for a relatively wide range of metallicity the parameters of the model can slightly adjust to accumulate a similar amount of metals. At higher spectral resolution and with adequate statistics, the degeneracy completely breaks and metal column densities are directly measured from absorption lines. 
	At redshift $z\gtrsim 1.5$, H and He absorption is no longer detectable in X-spectra of present X-ray telescope for the typical column densities observed in GRB environments, and the actual observed X-ray absorption is actually dominated by metals, with column densities comparable to those at lower redshift, if not higher (\cite{Campana+2012}).    
	
	Finally we stress that, as discussed in Paper I, the XMM sample is not biased toward luminous GRBs and covers a range of column densities consistent with those observed in the full population  making this sample, while small, representative of the broader GRB population. We thus conclude that the expected range of metal line intensities presented in Fig.\ref{fig:EWrest} could be considered reasonably representative of the environment of the population of (long) GRB.
	
	\section{Narrow X-ray lines in nearby GRB afterglows with current high-resolution spectroscopy missions}
	
	\subsection{XMM/RGS}\label{RGS}
	In Paper I we selected the 6 GRB afterglows with the highest fluence observed by XMM. We find  that, by self-consistently modeling the XMM-PN spectra with TEPID, the intrinsic continuum is absorbed by   gas heavily ionized by the GRB with density, size and metallicity  fully consistent with those of star forming regions. Here we take the analysis of these data a crucial step further, exploiting the plethora of multi-ionization narrow absorption lines expected to be imprinted in the GRB afterglow  X-ray spectrum, by such multi-phase and relatively dense gas. Because of their weak EW, those lines are not detectable by a low-resolution detector such as the PN, but they might be in the RGS.
	We have thus carried out an in depth search in the RGS data.

	\begin{table*}[!ht]
		\centering
		\renewcommand{\arraystretch}{1.25}
		\caption{Best-fit parameters and errors obtained by fitting the combined RGS1+RGS2 spectra, all errors are at 1$\sigma$.}
		\begin{tabular}{ccccccccc}
			\hline\hline
			& \multicolumn{3}{c}{Neutral fit} & \multicolumn{4}{c}{TEPID fit} & \\
			\hline
			GRB &  $N_{H}$ & C-stat/DOF & AIC & $N_H$ & log$n_0$ & C-stat/DOF & AIC & $\Delta$AIC \\
			& 10$^{21} \mathrm{cm}^{-2}$ & & & 10$^{21} \mathrm{cm}^{-2}$ & cm$^{-3}$ & &  \\
			\hline
			060729 & 0.91$^{+0.2}_{-0.2}$ & 3585/3278 & 3591 & 10$^{+30}_{-2}$ & 2.9$^{+1.4}_{-0.1}$ & 3573/3277 & 3581 & 10 \\
			061121 & 5.8$^{+0.14}_{-0.13}$ & 3819/3284 & 3825 & 60$^{+20}_{-50}$ & 3.3$^{+0.4}_{-1.4}$ & 3818/3283 & 3826 & 1 \\
			080411 & 3.4$^{+0.1}_{-0.1}$ & 3690/3298 & 3696 & 400$^{+170}_{-300}$ & 4.0$^{+0.8}_{-2.3}$ & 3685/3297 & 3693 & 3 \\
			090618 & 2.3$^{+0.04}_{-0.04}$ & 3591/3318 & 3597 & 15$^{+6.0}_{-2.0}$ & 2.4$^{+0.3}_{-0.1}$ & 3576/3317 & 3584 & 13 \\
			120711A & 14.7$^{+3.0}_{-3.0}$ & 3719/3313 & 3725 & 400$^{+200}_{-300}$ & 4.5$^{+0.3}_{-1.3}$ & 3716/3312 & 3724 & 1 \\
			221009A [exp1] & 5.0$^{+1.0}_{-1.0}$ & 3734/3318 & 3740 & 120$^{+270}_{-90}$ & 3.4$_{-0.6}^{+1.0}$ & 3729/3317 & 3737 & 3 \\
			221009A [exp2] & 5.0$^{+1.3}_{-1.2}$ & 3745/3317 & 3751 & 130$^{+320}_{-120}$ & 3.6$^{+0.8}_{-1.8}$ & 3745/3316 & 3753 & -2\\
		\end{tabular}
		\label{tab:RGS}
	\end{table*}
	
	\subsubsection{RGS data reduction}
	We retrieved the corresponding RGS ODF files from the XMM-Newton Science Archive \footnote{\url{https://nxsa.esac.esa.int/}} for each GRB of our sample. These were then locally reprocessed using the XMM Science Analysis Software (SAS) v22.1.0. We first processed the data using ~\texttt{rgsproc}, after which we extracted and checked the background lightcurve for each exposure to flag periods of background flaring. After filtering for any excess flaring above a count-rate of 0.2 counts s$^{-1}$, we then combined the RGS1 and RGS2 first order spectra using the task \texttt{rgscombine}. These are used in the subsequent analysis (an observation log is included in the Appendix Tab. \ref{tab:XMMlog}). We fit the spectra with the XSPEC package \citep{1996ASPC..101...17A} using Cash statistics \citep{1979ApJ...228..939C}, to account for the limited number of counts per channel (on average 1 count per channel) and error estimation is done with a 5$\times10^5$ step MCMC chain with a 1$\times10^5$ step burn-in. 
	
	\subsubsection{Narrow lines with TEPID}
	Consistently with Paper I, we compare the fits to the the RGS spectra with an absorbed  power law using a neutral ({\sc TBabs} \cite{2000ApJ...542..914W} model in XSPEC) and a TEPID absorber at the redshift of the burst, in addition to the absorption by our Galaxy. In this first run the TEPID abundances have been fixed to the solar value, consistently with the results obtained in Paper I. The results are presented in Tab.\ref{tab:RGS}. It is worth emphasizing that the TEPID parameters are consistent with those obtained by fitting the XMM-PN continuum, and that in basically all cases this model improves the fit compared to the cold absorber.

	To assess the statistical significance of the improvement, we have adopted the Akaike Information Criterion (AIC). For a given model with number of
	parameters $\theta$, the AIC can be estimated as:
	AIC = C-stat + 2$\theta$. When comparing two model fits to a given dataset, the model that gives the lower value of AIC is preferred over the other. The significance of the improvement in the fit can be numerically quantified by the difference $\Delta AIC$ = AIC$_{neutral}$ - AIC$_{TEPID}$. According to \cite{Burnham2004} and \cite{Fabozzi2014TheBO}, $\Delta AIC \ge 10$ would show conclusive support for TEPID over the neutral model, $\Delta AIC$ between 4 and 7 would indicate moderate support for TEPID over the neutral model, and $\Delta AIC \le  2$ would indicate no support for TEPID over the neutral model. The TEPID absorber is strongly preferred compared to neutral for GRB 090618 ($\Delta AIC$=13) and GRB 060729 ($\Delta AIC$=10), while in the other cases is either marginal or not significant. 
	
\begin{figure}
		\centering
		\includegraphics[width=0.95\linewidth]{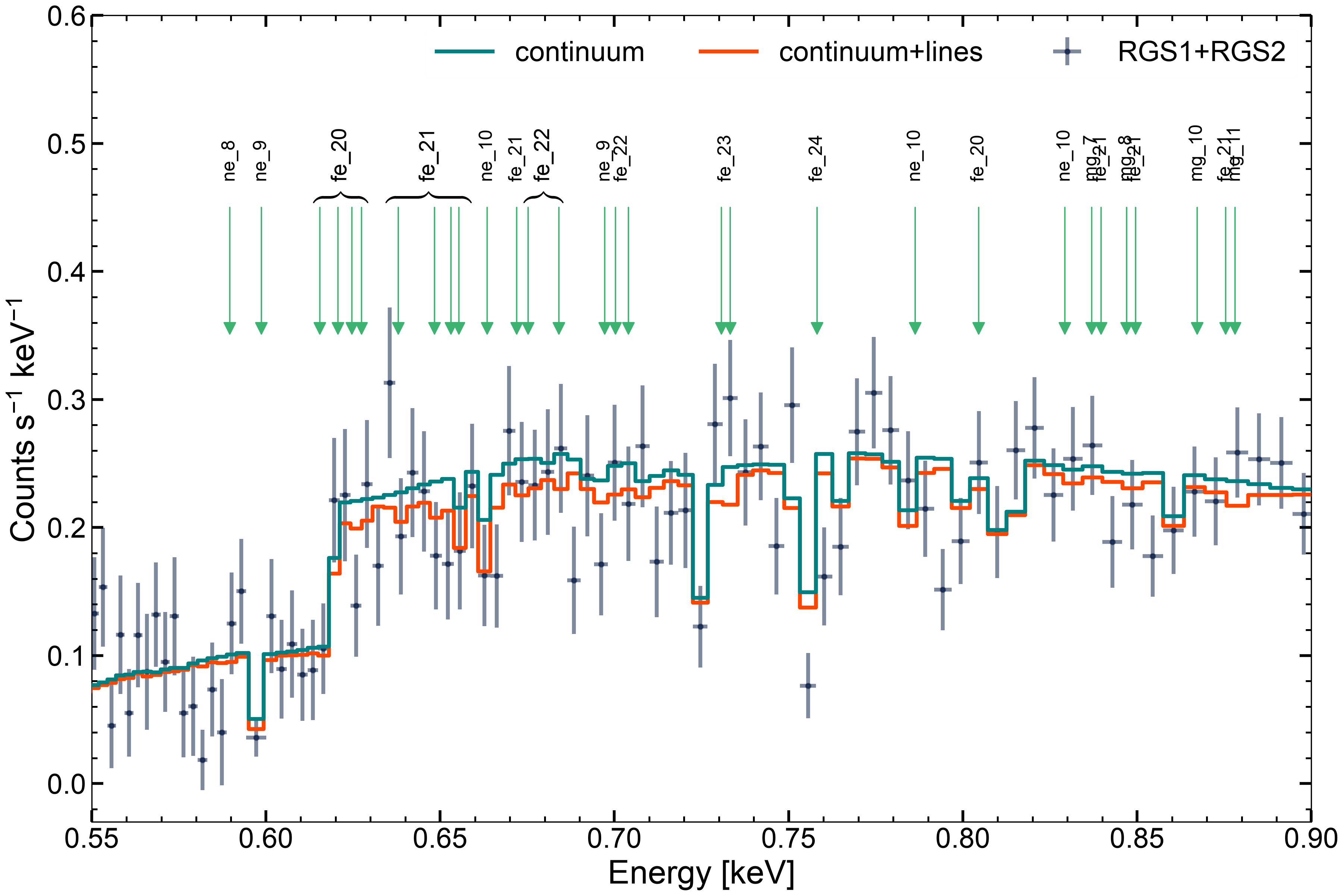}
		\caption{RGS1+2 combined count spectrum of GRB 090618 in the 0.55-0.90 keV range. The green solid line shows the best-fit continuum, while the red solid line includes the lines. The most prominent are identified with arrows, which shows that the improvement in the fit is driven by the combined detection of multiple transitions. A plot for the full fitted range of the RGS (0.33-2.5 keV) is included in the Appendix (Fig. \ref{fig:RGSfull090618}). Note that the dips present in the continuum are instrumental features.}
		\label{fig:GRB090618_lines}
	\end{figure}

	In order to ascertain to which extent the improvement of the TEPID absorber observed in GRB 090618 and GRB 060729 is due to the presence of absorption narrow lines as opposed to the photoionized continuum (absorption edges), we have added to the cold absorber model (our null model) the 10 most prominent absorption lines of the TEPID absorber, with EW ranging from 1 to 3 eV in the case of GRB 090618, and about 2-3 times weaker in GRB 060729. The relative intensities  of the lines have been fixed as well as the redshift, while their global normalization (relative to the value expected by the TEPID best fit model with solar abundance)  has been left free. In the case of GRB 090618 the fit finds a relative normalization  $I_{Lines}=0.9\pm0.3$, with an improvement $\Delta C= 7.5$. To assess the statistical significance of this improvement we have carried out 10,000 realizations of the null model, each one fitted with the null model without and with the addition of the lines. From the distribution of the resulting $\Delta C$ we derive a probability of $1.5\ 10^{-3}$ that the addition of the lines is due to a chance fluctuation, corresponding to 3$\sigma$ confidence level. We emphasize that the evidence is driven by the joint effect of several lines rather than a single one, the most prominent being in the 0.55-0.80 keV from NeIX, NeX and the L-shell lines from FeXX to FeXXII (Fig. \ref{fig:GRB090618_lines})
	
\begin{figure}
		\centering
			\includegraphics[width=0.85\linewidth, angle=90]{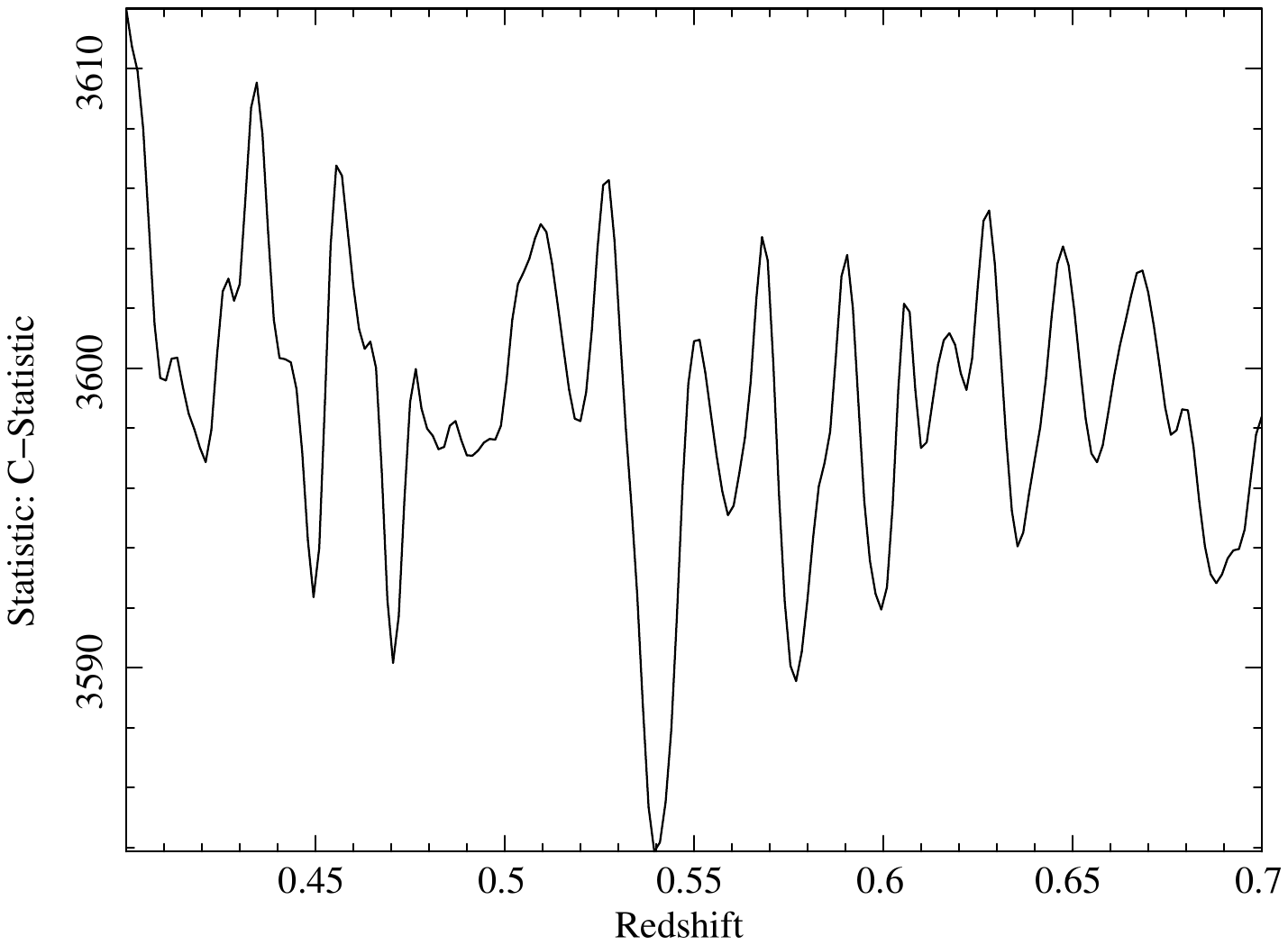}
		\caption{Scan of Cash statistics in redshift. A single deep minimum is identified at the optical redshit of GRB090618}
		\label{fig:z_scan}
	\end{figure}

\begin{figure}
		\centering
		\includegraphics[width=0.95\linewidth]{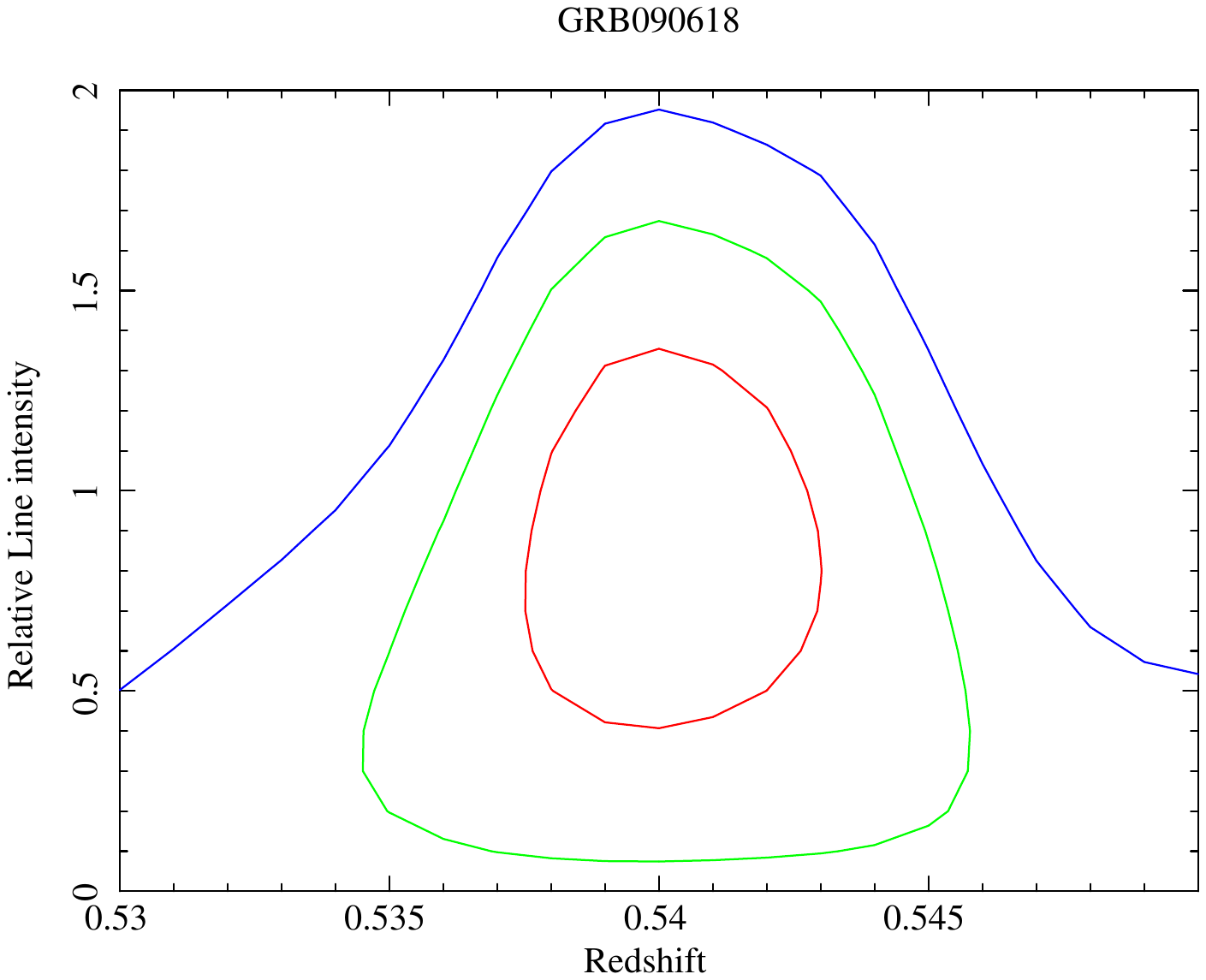}
		\caption{Contour plot of the Redshift vs Line intensity relative to the TEPID intensity at solar abundances. The solid lines correspond to 1,2 and 3 $\sigma$ levels for 2 parameters of interest}
		\label{fig:z_I_contour}
	\end{figure}

	To further explore the solidity of this result we have left the redshift free to vary. If the lines were merely due to continuum fluctuations, one should expect other  minima at the same depth, contrary to what we find (Fig.\ref{fig:z_scan}). There is only a minimum at $z=0.540\pm0.003$, fully consistent with the optical redshift. In Fig.\ref{fig:z_I_contour} we zoom in the best fit region, displaying the contour plot of the line intensity relative to that expected by the TEPID absorber with solar abundance. Although the line intensity is consistent with zero at $\lesssim 3 \sigma$  this does not contradict the evidence of a $3\sigma$ {\it detection.} They address two different questions. The detection significance  addresses the question: what is the probability that the detection is actually a fluctuation of a null model with no lines. On the other hand, the error of the intensity  addresses the question: what is the probability that a real line at a given intensity, when measured in a number of realizations, could result in a measurement of intensity consistent with zero, due to statistical fluctuations.

	Finally we note that the normalization of lines is proportional to the metal column density. In order to derive the metallicity a measurement of  H/He column densities is required. As show in Paper I, this is marginally possible  for low redshift burst such as this one. We have thus explored the TEPID model, scanning the metallicity in the range 0.05 to 2, finding a lower boundary (90\% level) $Z_\odot\gtrsim 0.1$.
	
	We have then considered GRB 060729, the other burst that displays a strong preference for  the TEPID model, as opposed to the cold absorber. We have adopted the same procedure followed for GRB 090618, including the 10 most prominent lines expected by the TEPID model. In this case the addition of the lines does not yield any significative improvement of the fit. This result is not surprising, because the line equivalent width expected from the TEPID absorber are two-three times weaker than in the case of GRB 090618. We conclude that, for this burst, the improvement exhibited by the TEPID model in fitting the RGS spectra is dominated by the absorption edges of the ionized absorber.
	
	\begin{figure}
		\centering
		\includegraphics[width=0.5\textwidth]{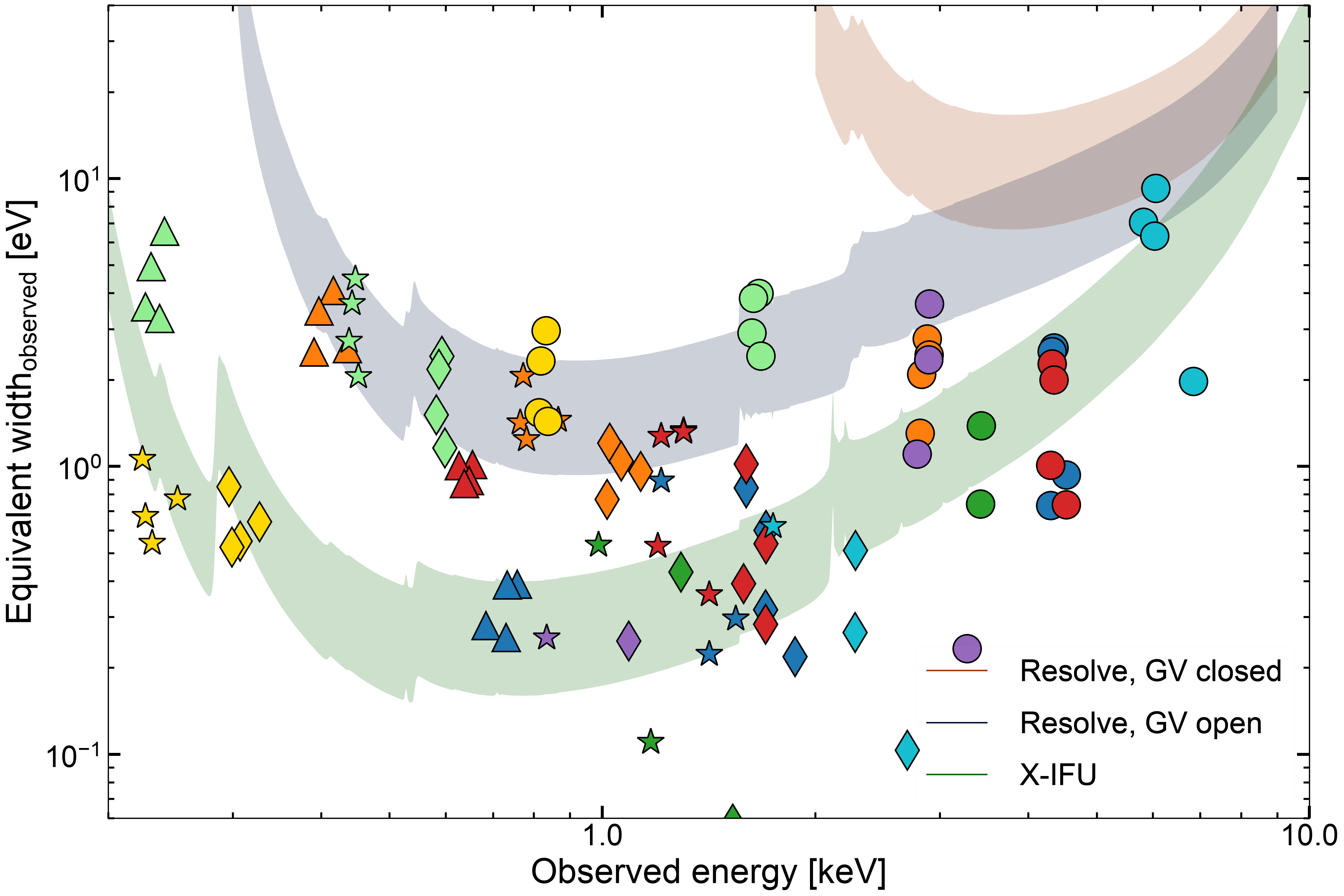}
		\caption{Minimum detectable equivalent width as a function of energy for X-IFU (green shaded region) and for Resolve with the gate valve open (blue shaded region) and closed (red shaded region). All three regions use an observed fluence of 5 $\times 10^{-7}$ and bracket a 5 to 2 $\sigma$ detection level. The spectral model for the minimum detectable equivalent widths shown in the graph assumes a powerlaw of index 2 with a Galactic column of 3$\times 10^{20}$ cm$^{-2}$ and an intrinsic column of $5 \times 10^{21}$ cm$^{-2}$ at a redshift $z = 2$. Overplotted, are the strongest absorption lines predicted for silicon (stars), sulfur (diamonds), iron L-shell (triangles) and iron K-shell (circles), color-coded per GRB: GRB 060729 (dark blue), GRB 061121 (orange), GRB 080411 (green), GRB 090618 (red), GRB 120711A (magenta), GRB 221009A (cyan), along with a medium bright GRB at z=3 (light green) and z=7 (gold).}
		\label{fig:EW_XIFU_XRISM}
	\end{figure}
	
	\subsection {\textit{XRISM} and the closeby GRBs}\label{XRISM}
	By adopting the XMM sample as reference we have  explored the possibility to detect ionized absorption lines with \textit{XRISM}/Resolve \citep{2021SPIE11444E..22T}. In Fig.\ref{fig:EW_XIFU_XRISM} we plot for each burst and for the 4 most prominent  lines of Fe (both K and L shell), Si and S, the EW expected by the TEPID absorber at the observed energies. 
	As reference we plot the narrow line sensitivity of Resolve for a relatively bright fluence that is feasible in a \textit{XRISM} observation  (see Sect. \ref{Sect:fluence}) $\mathcal{F}=\ 0.5\times 10^{-6}$~ erg~$\mathrm{cm^{-2}}$ for both the closed and open gate valve configurations. Although each individual line falls below the 5 $\sigma$ sensitivity, it is the joint contribution that needs to be considered. For example, let us consider the Fe lines expected for GRB 221009A. The three strongest lines summed up to a total EW=25 eV, yielding a total S/N=3.6, having corrected the sensitivity by the factor $\sqrt{3}$. For the closed configuration, where the sensitivity degrades below about 3 keV, the only line that can be detected is therefore the K- Fe for $z\lesssim1$.
	The perspectives improve significantly, should the gate valve be successfully open. In such case, lines from other elements would become accessible, also for a wider range of  redshift. Individual lines from  Fe, Si,   are  close to the 5$\sigma$ detection threshold.
	
	\begin{figure}
		\centering
		\includegraphics[width=0.5\textwidth]{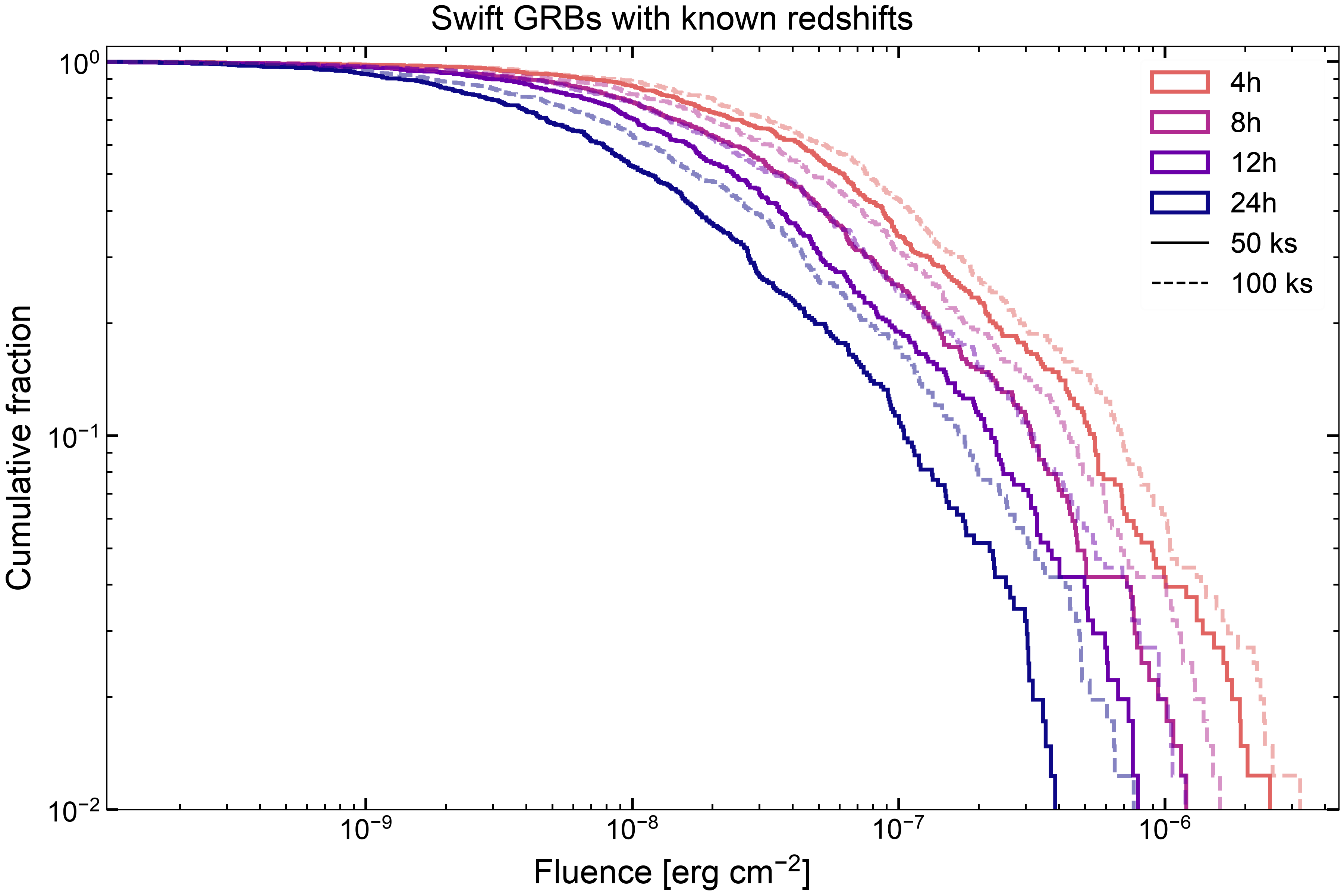}
		\caption{Cumulative distribution function of the observed fluences in the \textit{Swift} population of GRBs with known redshifts. From lighter to darker colors, we estimate the fluence starting from 4h post-GRB upto 24h post-GRB. The solid line shows the fluence obtained in a 50 ksec exposure, instead the dashed line show the fluence obtained in a 100 ksec exposure.}
		\label{fig:GRBcdf}
	\end{figure}

	\subsubsection{ Fluence  distribution  of GRB afterglows}
	\label{Sect:fluence}
	In order to assess the feasibility of a follow-up observation of a GRB afterglow yielding a given fluence, we have analysed the full set of afterglow light curves produced by \textit{Swift} and computed the integral distribution function of the afterglow fluence for  selected set  of the start time (from 4 to 24 hours) and observation duration (50 and 100 ksec). In Fig.\ref{fig:GRBcdf} we show that, for a TOO observation starting 1 day after the GRB and lasting 100 ksec, a realistic observational strategy with XRISM,  the fraction of afterglows with  $\mathcal{F}\gtrsim 0.5\times 10^{-6}$~ erg~$\mathrm{cm^{-2}}$ is 0.02.
	Taking into account the \textit{XRISM} field of regard (1/2 sky) and the number of GRB localized by currently operated GRB facilities ($\gtrsim 300$ per year), one expects that about 3 GRB afterglows with such fluence be observable by XRISM/Resolve per year.
	We conclude that there are solid prospects to detect narrow lines from ionized absorbers should a TOO program on bright afterglows by \textit{XRISM} be carried out.
	
	\section {From popII to popIII environments  with NewAthena}\label{NewAth}
	
	Thanks to the combination of the high energy resolution and low background enabled by the non-dispersive detector of the XIFU microcalorimenter \citep{2023ExA....55..373B, 2025ExA....59...18P} and the large collecting area of the mirrors, \textit{NewAthena} will deliver an exquisite sensitivity to weak lines. We show that it could routinely provide high resolution spectra for a vast number of GRB afterglows, only limited by the time dedicated to such an observing program. Fig.\ref{fig:EW_XIFU_XRISM} shows that most of the lines predicted by TEPID in the XMM sample should be detectable by XIFU.
	
	\begin{figure}
		\centering
		\includegraphics[width=\linewidth]{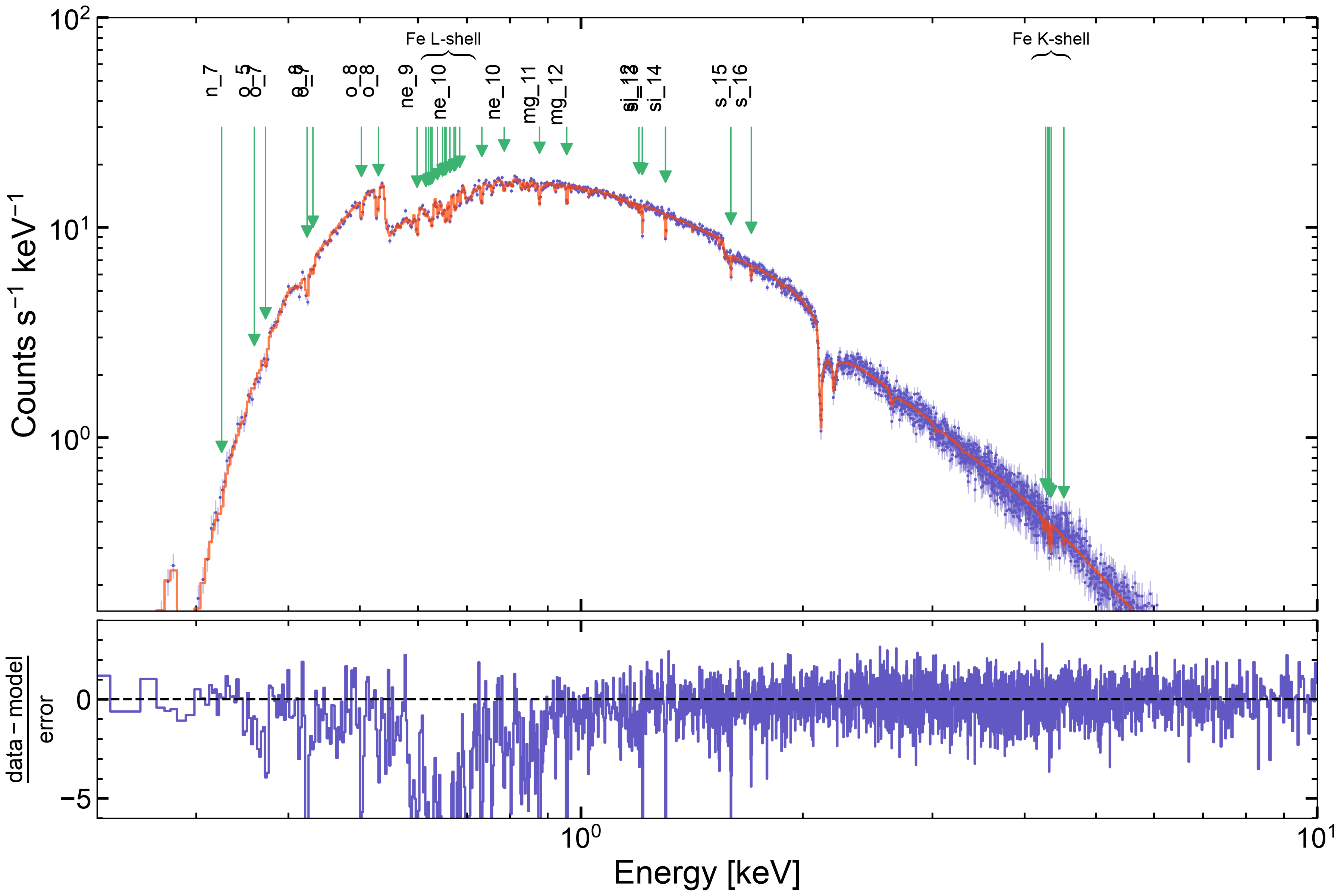}
		\caption{XIFU simulated spectrum for GRB 090618, assuming the best-fit parameters obtained from the XMM-PN fit. The spectrum has been dynamically rebinned to improve the visibility of individual line features. Individual lines having an observed equivalent width higher than 0.5 eV are marked with vertical dashed green lines. The bottom panel plots the residuals of the lines against the continuum.}
		\label{fig:090618}
	\end{figure}

	To assess the improvement more quantitatively, we have sim-
	ulated the XMM sample of GRB afterglows as they would be
	observed by the XIFU instrument. We have adopted the latest version of the response matrices and effective area \citep{2025ExA....59...18P}. To allow for a fair comparison, we have assumed the same TOO start time of XMM for each burst (ranging from 5 to 22 hours, with the exception of GRB221009A, where the observations were carried out at about 2 and 4 days after the burst, see Tab..\ref{tab:XMMlog}). These times bracket  the conservative requirement of \textit{NewAthena} to start a TOO observation 12 hrs after the trigger. For the integration time we adopted the \textit{NewAthena}  baseline of 50 ksec XIFU exposure during TOOs. The spectra have been simulated using the \textit{fakeit} command in XSPEC assuming the XMM-PN best fit TEPID parameters. We then refitted the spectra to gauge the capability to measure the key parameters.
	
\begin{figure}
		\centering
		\includegraphics[width=\linewidth]{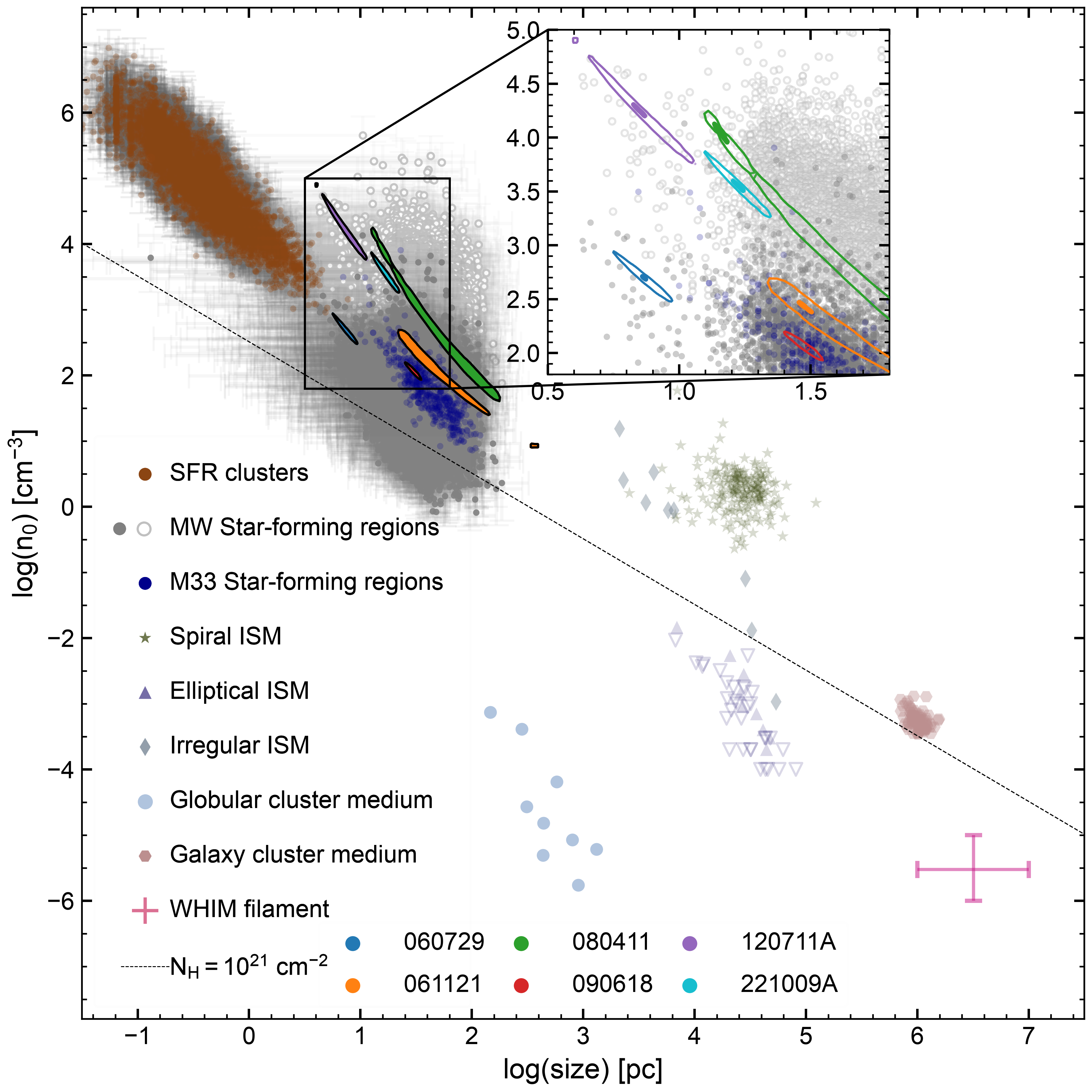}
		\caption{Contours of size vs density estimated by fitting the XMM-PN spectra of the Paper I sample of GRBs. Plotted for comparison are the densities and sizes of gravitationally bound structures in the Universe in increasing order of size: Clusters inside star-forming regions (brown); Galactic (filled (uniform dense gas filling factor) and open grey points (dense gas filling factor of 0.3) and extragalactic (filled blue points) star-forming regions; globular cluster medium (light blue circles); the interstellar medium of spiral (green stars); irregular (blue-gray diamonds) and elliptical galaxies (purple triangles); galaxy cluster medium (red hexagons) and a WHIM filament (red cross). In the inset panel, the filled contours demonstrate the vast improvement obtained by fitting the X-IFU spectrum compared to the unfilled XMM-PN contours. All contours are plotted at 90\%.}
		\label{fig:contours}
	\end{figure}

	In all cases, we rebinned the XIFU simulated spectra to four channels and use Cash statistics to fit the spectra. Error estimations are done with a 5$\times10^5$ step MCMC chain with a 1$\times10^5$ step burn-in. The rich spectrum of lines imprinted on the XIFU spectra (Fig. \ref{fig:090618}) allows us to determine the physical and chemical parameters of the absorber with exquisite precision. The first key comparison is on the density and the size of the ionized absorber that, as pointed out in Paper I, are pinpointing star forming regions embedding the GRB. In Fig. \ref{fig:contours} we compare the vast improvement in measuring these  parameters that, as noted in Paper I, are fundamentally characterizing the close, star forming, environment of the GRB. We have included in the Appendix the contour plots for the XIFU fits of GRB 090618 and GRB 080411, as they bracket the typical spread in errors retrieved from our fits. (Figs. \ref{fig:0906_fixed}, \ref{fig:0804_fixed})
	
	The second result is the capability of determining the redshift with a relative error $\lesssim 10^{-3}$ (Tab. \ref{tab:summary}). This opens the possibility to get the redshift directly from the X-ray afterglow with \textit{NewAthena}, without resorting to an optical or IR counterpart to be discovered. This is particularly crucial when one considers that only about 50\% GRBs have optical counterparts while more than 90\% have an X-ray afterglow. The lack of optical afterglow in the so-called {\it dark} GRBs is likely a combination of different causes \citep{Piro2002,Jakobsson+04}, the prevailing ones being dust extinction  and lack of fast and deep enough optical follow up. It is also possible that some fraction of optically dark GRB are high-z burst, in which the optical flux is damped by the Ly${\alpha}$ forest. 
	
	The possibility of measuring {\it metallicity} by X-ray absorption in the 0.2-10 keV range with \textit{NewAthena} is effectively enabled by the presence of absorption from H and He. It has become clear (see \citealp{Watson+10} \& Paper I) that, even in the presence of a strong ionization flux from close to the GRB, the vast number density of H and He allows a significant fraction of them to survive the ionization and be effective in absorbing X-ray with an opacity that can be comparable to that of metals upto (rest-frame) energies of about 1-2 keV. It could then be possible  to measure with X-IFU both the effect of the lineless absorption imprinted by H and He, and the total column density of metals. The latter is pinned down by exploiting not only their effect on the continuum by edges but, foremost, the lines. 
	
	We have thus fitted again the simulated spectra leaving the H and He column densities, on one side, and the metals, on the other side, as free parameters (normalized to the input quantities). We then computed the metallicity (normalised to solar) as the ratio between them (including all samples from the MCMC chain), obtaining typical errors $\lesssim10\%$ for all but one of the closer GRBs, while for the GRB at z=3 the error is still an acceptable $\approx 20\%$ (Tab. \ref{tab:summary}, see also Fig. \ref{fig:0906_varmet}-\ref{fig:metz3}). In one extreme case, namely GRB 080411, the total ionizing energy is very high, ($\approx 10^{55}~$erg in the 0.3-10 keV range, see Tab. \ref{tab:XMMlog}). It is 5 times larger than the second (GRB 221009A, i.e. the BOAT, that still predicts strong lines) and $\gtrsim 10-20$ times larger than the others. In such extreme ionization condition, the metallicity is not well determined, since H and He are almost completely ionised and, thus, their columns cannot be constrained accurately. The presence of fainter metal lines let the fit include a wider range of allowable ionization, giving in turn a larger error in the ionic fractions of H and He ions. However, as the ionizing flux is a direct observable quantity, one can straightforwardly identify any of such extreme cases. Finally, for the burst at z=7 the metallicity is not constrained. While the metal column density can be measured with good accuracy (Tab. \ref{tab:summary}), that is not the case for H (and He) absorption, that shifts to the lower limit of the X-ray bandwidth.
	
	Finally, the sensitivity to lines of different elements suggests that individual abundances of the most prominent metals can be obtained. We then leave free the individual column densities of Si, S and Fe (normalised relative to the input quantities) as a test. In Tab. \ref{tab:summary}, the original columns are recovered with errors on the order of $\approx 10-20\%$ (see also Fig. \ref{fig:0906_fracs}). The errors on the individual elemental columns are found to be inversely proportional to the number of lines from the ions of that element. This is clearly seen for GRB 090618, which has multiple lines of Si, S and Fe (see Fig. \ref{fig:090618}), and thus the relative errors on the columns are around 10\%. Instead, the error is correspondingly larger for Si and S in the case of GRB 120711A, as for these elements fewer lines are being fitted, while Fe is still relatively well constrained due to the presence of multiple Fe-K lines. 
	
	\subsection{Population study of the close environment across the Cosmic Ages }
	
	Herewith we discuss to which extent the results presented above on the XMM sample can be obtained on a larger sample of GRB, thus enabling population studies.
	For the nominal 12 hrs TOO, and the fluence encompassed by the Paper I sample, the corresponding cumulative fractions range from $\approx 0.1$ for mid-bright fluences  $ 0.2-0.3\times 10^{-6}$~ erg~$\mathrm{cm^{-2}}$, to  $\approx0.01$ for the very bright fluence  $ \approx 1\times 10^{-6}$~ erg~$\mathrm{cm^{-2}}$ (Fig.\ref{fig:GRBcdf}). We note that this sample extends to z$\approx1.5$. 
	
	We have thus explored the capability to carry out similar measurements for somewhat fainter, thus even more common bursts, with a fluence $= 0.1\times 10^{-6}$~ erg~$\mathrm{cm^{-2}}$, corresponding to a cumulative fraction of 0.2. We have adopted the average best fit parameters of the absorber as derived from the XMM sample, log n = 3.0 and log N$_H$=22.8, with solar metallicity, and run the corresponding TEPID computation. In  order to extend the redshift range probed by the XMM sample, we have selected z=3 and z=7. The ionizing luminosity of the burst in the rest frame has been normalized so to yield the observed fluence. As for the fluence adopted for the high redshift burst we point out that our fiducial value is realistic if not conservative: of the existing sample of high redshift bursts at $z\gtrsim 6$, at least two have  afterglows with comparable fluence under the same conditions of a \textit{NewAthena} TOO: 12 hrs start with a 50 ksec observation. In passing we note that for a TOO initiated 4 hrs after the burst, the fluence would double.
	
\begin{figure}
		\centering
		\includegraphics[width=\linewidth]{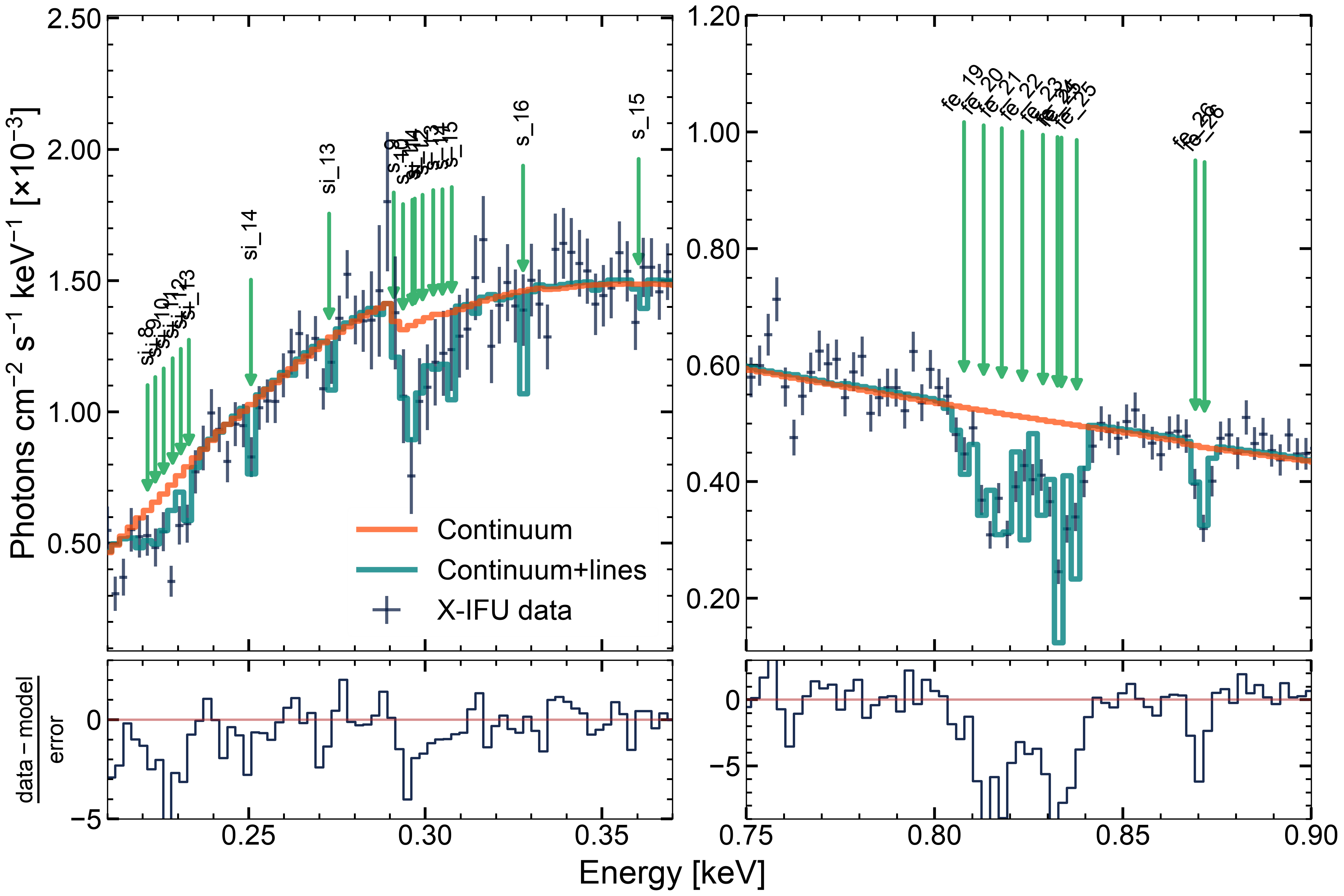}
		\caption{X-IFU simulation for a $z=7$ GRB with an observed fluence of 1e-7 cgs assuming a ToO at 12 hr with 50 ksec of integration, zooming in on the Si complex (left panel) and the Fe-K shell complex (right panel). The top panel plots the unfolded spectrum (with a bin size of 1 eV) in navy blue. The continuum is plotted in orange, while the total spectrum is plotted in green. The bottom panels plot the residuals for the lines against the continuum.}
		\label{fig:z7PopIIGRB}
	\end{figure}

	The simulated spectrum has then been refitted with the same procedure as the Paper I sample. The results are presented in Tab.\ref{tab:summary}. In summary, even for fainter, more distant bursts, up to z=7, the key parameters of the close environment (density, size) and the redshift of the GRB can be derived with good precision (see also Fig. \ref{fig:z7_fixed}). Metallicity can be determined up to z$\approx3$ (Fig. \ref{fig:metz3}), while the column density of prominent metal elements (hence) the relative error on abundances, depends on the element. In Fig.\ref{fig:z7PopIIGRB} we show the spectrum of the z=7 burst exhibiting strong Fe-K line complex, with lower energy line complex from Si and S lines. Iron abundance can be measured with $\lesssim 10\%$ error upto z=7, thanks to the strong Fe-K line complex at $\approx 7~$keV rest-frame, while for Si and S the relative error is $\approx 30-40\%$ (see also Fig. \ref{fig:z7_fracs}). We note that the relative error on abundance is larger than the line intensity because it includes the error on the ion fractional distribution estimated from modelling.
	
	\subsection{Rate of GRB observable by NewAthena}
	
	To assess the rate of GRB that can be observed by \textit{NewAthena} we have taken into account the  \textit{Swift} GRB rate ($\approx 80\ {\rm yr}^{-1}$, \cite{Lien2016}), the field of view of its BAT instrument, 1.4 sr, and the downtime due to SAA passage \citep[$\approx 20\%$][]{Barthelmy2005, Burns2016}. We thus derive the number of GRB afterglows above a given fluence that can be observed by \textit{NewAthena} as: $N_{GRB(>\mathcal{F}_0)}\approx10^3\times \eta(>\mathcal{F}_0)\times\eta_{FOR}\times\eta_{GRBM}~\mathrm{yr}^{-1}$, where $\eta(>\mathcal{F}_0$)  is the fraction of afterglows above $\mathcal{F}_0$ derived from Fig.\ref{fig:GRBcdf}, $\eta_{FOR}=0.34$ is the field of regard, i.e. the fraction of the sky accessible to \textit{NewAthena} telescope at any given time  \citep{2025NatAs...9...36C}, and $\eta_{GRBM}$ is fraction of the sky covered by GRB instrument(s). Taking the fraction of afterglows above the given fluence with Tstart=12 hrs and a 50ks exposure (Fig.\ref{fig:GRBcdf}) we derive that \textit{NewAthena} can observe upto 50 GRB afterglow per year with $\mathcal{F}\gtrsim 0.1\times 10^{-6}$~ erg~$\mathrm{cm^{-2}}$ and upto 10 bright afterglows per year with $\mathcal{F}\gtrsim 0.5\times 10^{-6}$~ erg~$\mathrm{cm^{-2}}$. This assumes a reasonable scenario where $\approx 70\%$ of the sky is covered by one or more GRB experiments involving  e.g. constellations of nano-satellites, ESA F-proposal experiments like GRINTA \citep{Rodi2025}, a successor of Einstein Probe \citep{yuan2025science} or SVOM \citep{cordier2026svom}. The number of dedicated TOO, at most 4 per month, is consistent with the \textit{NewAthena} capability. We conclude that a dedicated program of follow-up observation can lead to a sample of upto 200 afterglows in 4 years, with a total investment of 10 Ms in observing time.

	\begin{figure}
		\centering
		\includegraphics[width=0.5\textwidth]{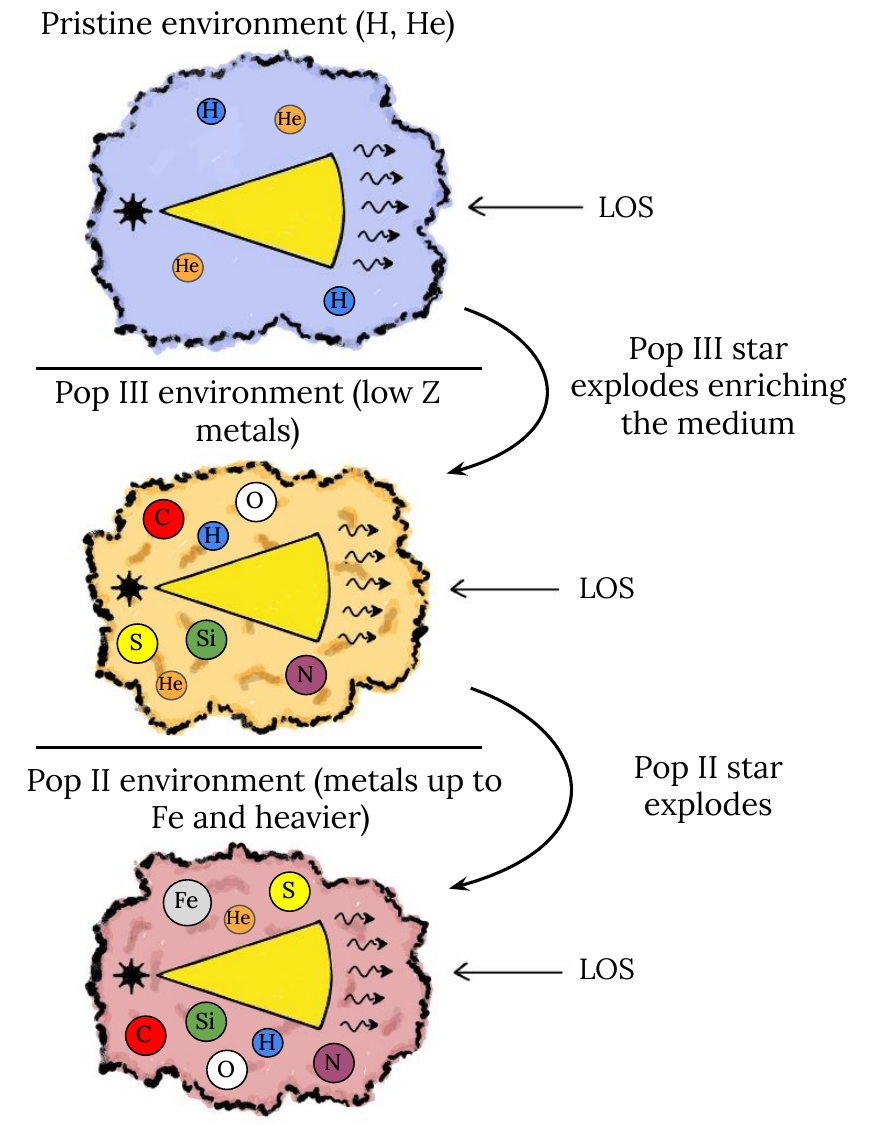}
		\caption{Sketch of the various close environments that can be probed by X-ray absorption spectroscopy  at high redshift. Upper panel: GRB exploding in a pristine medium. Mid panel: a GRB exploding in a medium enriched by popIII ejected metals upto Si and S. Lower panel: a GRB exploding in a popII environment, including Fe and heavier metals }
		\label{fig:highzenvirons}
	\end{figure}
	
	\begin{figure*}[h]
		\centering
		\includegraphics[width=0.8\linewidth]{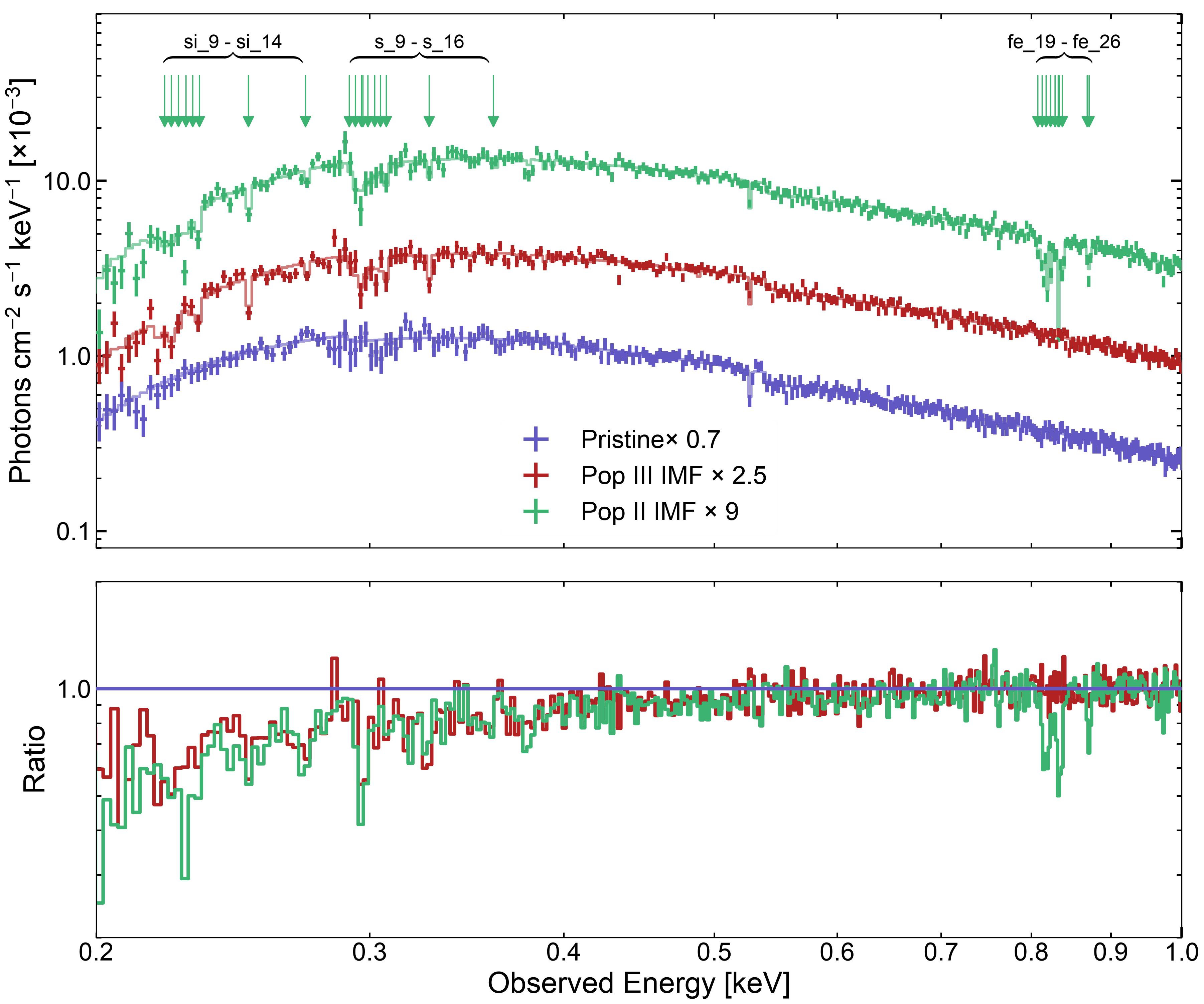}
		\caption{X-IFU unfolded spectrum for three distinct progenitor environments: pristine (violet: no metals), popII medium (green: Si and Fe) and a medium from popIII with  top-heavy IMF (red: Si, but no Fe). The bottom panel plots the ratio of the counts normalised to the pristine case. For clarity of presentation, we have zoomed in on the 0.2-1 keV region. The absence of Fe-K absorption lines around 0.8 keV in the red spectrum as well as the absence of any lines in the pristine spectrum can be clearly seen. Note the narrow feature at 0.53 keV and the residual curvature at low energy of the pristine medium is produced by absorption by our Galaxy.}
		\label{fig:compare_spec}
	\end{figure*}
	\begin{figure*}
		\centering
		\includegraphics[width=0.8\linewidth]{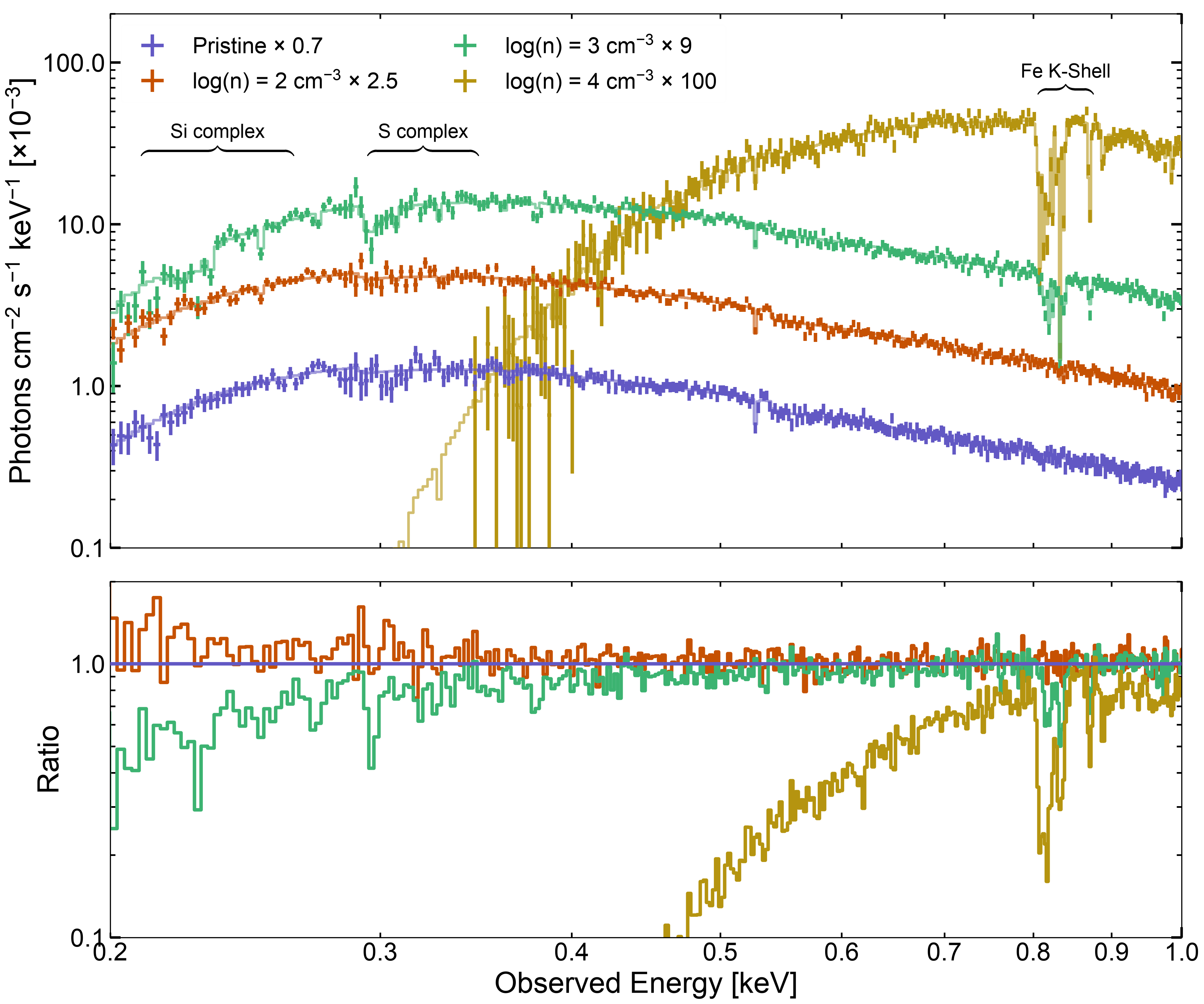}
		\caption{X-IFU unfolded spectrum for a pristine environment (violet: no metals) and a popII environment with three distinct number densities: log(n) = 2 cm$^{-3}$ (dark orange), log(n) = 3 cm$^{-3}$ (green) and log(n) = 4 cm$^{-3}$ (dark yellow). The spectra for the popII environment are simulated by fixing the size at 20 pc, thereby scaling the column with the number density. The bottom panel plots the ratio of the counts normalised to the pristine case. For clarity of presentation, we have zoomed in on the 0.2-1 keV region. The ratio for log(n) = 2 cm$^{-3}$ does not show prominent narrow features, showing the limit of sensitivity for distinguishing a genuine pristine environment against a ionised popII medium. Note the narrow feature at 0.53 keV and the residual curvature at low energy of the pristine medium is produced by absorption by our Galaxy.}
		\label{fig:PopIImimic}
	\end{figure*}
	\subsection{Telling popII from popIII environment?}
	Population III (or Pop. III) stars, the first stellar generation built  from metal-free primordial gas, first started to form at redshifts $z\gtrsim20$ \cite[for a review]{Glover_popIII_26}. The Initial Mass Function (IMF) of Pop III stars is currently modeled as top-heavy, with most of the mass above 10-20 M$_{\odot}$. With such large masses the popIII star evolves rapidly, in a few million years, finally  exploding. Stars with masses between 10 and 40 M$_{\odot}$ explode via a core-collapse supernovae \citep{Heger+2003}, possibly producing a GRB powered by the newly formed black hole \citep{Toma+2016}. The supernova ejecta enrich the close environment with the very first  metals synthesized by the star, and thus lay the chemical foundations for the birth of subsequent generation of (Pop II) stars. Probing popIII {\it environments} does not necessarily require a popIII-GRB as a probe. While there are solid theoretical prediction as to their existence \citep{Suwa+2011,Nakauchi+2012}, those explosions are yet to be identified in the present population of long GRBs with a measured redshift, that are associated to popII massive stars. Because the chemical enrichment is highly inhomogeneous, popIII and popII stars coexists for a long period, to the extent that popIII-enriched environments can still be present at redshift $z\lesssim5$ \citep{Tornatore+2005,Jaacks+2019}: popII GRBs can thus explode in  popIII environment. 
	
	We have therefore considered three cases (Fig.\ref{fig:highzenvirons}): a) a pristine environment, made of H and He without metals; b) a region enriched by top-heavy popIII supernovae and compared them with the c) popII-environment described in the previous subsection. In order to provide a controlled comparison, we have assumed the same fiducial values of column density and density (hence the size) and the same fluence of the popII GRB at high-z. The corresponding spectra as observed by XIFU are shown in Fig.\ref{fig:compare_spec}.
	
	In the case of a pristine medium, the absorption is only due to H and He: we  do not expect metal features but only a mild curvature below 0.3-0.4 keV, because H and He absorption at z$\approx 7$ shifts close to the lower end of of the X-ray band. The spectrum is close to a pure power law just absorbed by our Galaxy. These properties are readily apparent when we compare the corresponding simulated spectrum with those obtained in the cases of a metal enriched media (Fig.\ref{fig:compare_spec}), exhibiting both narrow lines and continuum absorption. On the other hand, a popII environment with a low column density could potentially mimic a pristine environment, due to all metals (particularly Fe) being heavily ionised. To test the limiting case, we therefore checked for the average size of 20 pc, the size corresponding to the fiducial value of log(N$_H$) = 22.8 cm$^{-2}$ and log $n_0$=3 the effect of varying the number density of the medium (as a lower density then equates to a lower column). At number densities log$n_0$ = 3,4 cm$^{-3}$, the medium exhibits both strong lines and an absorption continuum, while  at log$n_0 \lesssim 2$, these lines are at the limit of the X-IFU sensitivity and the overall shape becomes undistinguishable from a pristine gas (Fig.\ref{fig:PopIImimic}).  On the other hand such low densities are unlikely to characterize star forming regions at high redshift. In this regime, it is likely to take place at much larger densities, log(n)$\gg 3$, so to enable high star forming efficiency required to explain  the bright high-z galaxies discovered by JWST  \citep[e.g.][]{Deckel+23,Chen+26}. 
	
	A popIII explosion is expected to inject in the close environment a global amount of metal  not unlike that of a popII core-collapsed supernova but with two distinct differences. The first is  a strong odd-even abundance pattern, where elements with odd atomic number have abundance  comparatively lower than elements with even number \citep{Heger+2003}. To take this effect into account, we have updated our TEPID code, as described in Sect. \ref{sec:TEPID}, including  particular heavy elements with odd atomic numbers, i.e  those that can potentially fall in the XIFU bandpass at high redshift. However the expected lines are extremely weak, because of  the combination of cross sections and abundances of odd elements. Some potentially detectable line could be produced from low-Z odd elements, but fall outside the bandwidth at such high redshift. 
	
	Another distinct feature that could characterize the medium enriched by popIII stars with a top-heavy IMF is the paucity of heavy elements. The extent of reduction depends on the explosion model and the mixing in the progenitor star, with an extreme reduction (order of magnitude or more) in particular for Fe \citep{Heger+2003}. The simulated spectrum  presented in Fig.\ref{fig:compare_spec} is particularly compelling, especially when compared to the popII case, for the lack of Fe lines while lines from lower Z elements, such as S and Si are clearly visible. In order to better quantify the capability of measuring the content of Fe vs Si or S, we have fitted the spectra (popII-enriched and popIII-enriched) leaving the corresponding abundances free to vary. In the latter case, the ratio of Fe over Si and S leads to an upper limit $\lesssim 0.15$, significantly below the unity ratio expected for a popII environment (see Figs. \ref{fig:z7_fracs}, \ref{fig:z7th_fracs}). 
	
	\begin{table*}[!ht]
		\centering
		\renewcommand{\arraystretch}{1.5}
		\caption{Best-fit parameters and errors obtained by refitting the simulated X-IFU spectrum; errors are reported at 1$\sigma$. The contour plots corresponding to the reported values are presented in the appendix A.3}
		\begin{tabular}{c c c c c c c c}
			\hline\hline
			GRB & $z$ & logNH & log$n_0$ & Metallicity & $A_{Si}$ & $A_{S}$ & $A_{Fe}$ \\
			\hline
			060729 & 0.54$^{+0.0001}_{-0.0001}$ & 21.96$^{+0.006}_{-0.001}$ & 2.77$_{-0.002}^{+0.01}$ & 1.1$^{+0.06}_{-0.06}$ & 1.0$^{+0.1}_{-0.1}$ & 1.0$^{+0.2}_{-0.1}$ & 1.1$^{+0.1}_{-0.1}$  \\
			061121 & 1.314$^{+0.0002}_{-0.0002}$ & 22.39$^{+0.004}_{-0.003}$ & 2.55$_{-0.005}^{+0.005}$ & 1.0$^{+0.04}_{-0.04}$ & 0.9$^{+0.05}_{-0.05}$ & 0.8$^{+0.10}_{-0.10}$ & 0.8$^{+0.02}_{-0.05}$  \\
			080411 & 1.028$_{-0.001}^{+0.001}$ & 23.60$^{+0.05}_{-0.02}$ & 4.00$^{+0.05}_{-0.10}$ & 0.9$^{+0.9}_{-0.4}$ & 1.0$^{+0.40}_{-0.30}$ & 1.0$^{+0.50}_{-0.40}$ & 1.0$^{+0.60}_{-0.40}$ \\
			090618 & 0.54$^{+0.0001}_{-0.0001}$ & 21.99$^{+0.003}_{-0.003}$ & 2.13$_{-0.01}^{+0.01}$ & 1.0$^{+0.05}_{-0.05}$ & 1.0$^{+0.10}_{-0.10}$ & 1.0$^{+0.10}_{-0.10}$ & 1.0$^{+0.03}_{-0.03}$\\
			120711A & 1.405$^{+0.0003}_{-0.0003}$ & 23.46$^{+0.02}_{-0.04}$ & 4.24$_{-0.03}^{+0.07}$ & 0.9$^{+0.10}_{-0.10}$ & 1.0$^{+0.40}_{-0.60}$ & 0.5$^{+0.40}_{-0.40}$ & 1.0$^{+0.10}_{-0.20}$ \\
			221009A & 0.151$^{+0.0001}_{-0.0001}$ & 23.15$^{+0.06}_{-0.01}$ & 3.51$_{-0.01}^{+0.09}$ & 0.9$^{+0.10}_{-0.10}$ & 1.0$^{+0.20}_{-0.30}$ & 1.0$^{+0.30}_{-0.30}$ & 1.0$^{+0.10}_{-0.10}$ \\
			PopII GRB (z=3) & 3.00$_{-0.001}^{+0.001}$  & 22.80$_{-0.01}^{+0.01}$  & 3.00$_{-0.05}^{+0.01}$  & 1.2$^{+0.20}_{-0.20}$ & 1.0$^{+0.10}_{-0.10}$ & 1.2$^{+0.40}_{-0.10}$ & 1.0$^{+0.07}_{-0.04}$  \\
			PopII GRB (z=7) & 7.00$^{+0.001}_{-0.001}$ & 22.80$^{+0.02}_{-0.01}$ & 3.03$_{-0.03}^{+0.06}$ & - & 1.0$^{+0.40}_{-0.30}$ & 0.9$^{+0.40}_{-0.30}$ & 1.0$^{+0.10}_{-0.10}$ \\
			PopIII-enriched GRB (z=7) & 7.00$_{-0.01}^{+0.01}$  & 22.60$_{-0.10}^{+0.20}$  & 2.51$_{-0.20}^{+0.50}$ & - & 1.4$^{+0.30}_{-0.60}$ & 0.7$^{+0.50}_{-0.20}$ & <0.06 \\
			\hline
		\end{tabular}
		\label{tab:summary}
	\end{table*}
	
	\section{Conclusions}\label{Conclusions}
	
	Absorption spectroscopy of GRB afterglows in the optical and infrared bands  provides crucial information on the properties of the host-galaxies of GRBs from the local to the high-z Universe \citep{Vreeswijk2007, Saccardi2026}. On the other hand, the  star forming region embedding the progenitors of GRBs on scales of $\lesssim 100$ pc remains inaccessible at these wavelengths, being heavily ionized by the GRB prompt and afterglow emission, to the extent that it  can only be probed by X-ray spectroscopy. Indeed X-rays spectra of afterglows display strong absorption, with column densities 10 times larger than those measured in the optical, in strong support of a close dense and ionized medium. X-ray absorption spectroscopy of GRB afterglows can thus potentially  provide a unique probe of the star-forming environment of GRBs across the Universe. However, two necessary and complementary conditions  are to be met. The first is a photoionization code that properly takes into account the strongly varying ionizing luminosity and the spatial stratification of the ionization of the medium close to the GRB. The second is the availability of space-based instruments capable of exploiting the powerful diagnostics offered by the rich narrow absorption features arising in this medium. In this paper we show that both conditions are met and that high resolution X-ray spectroscopy of GRB afterglows is becoming a concrete opportunity. 
	
	We use the TEPID code \citep{Luminari2023} to derive the ionization structure of the absorber and the expected narrow absorption lines. Such a spatially stratified, time-variable ionization imprints a rich and unique variety of narrow absorption lines spanning a ionization range much wider than single-zone, equilibrium photoionization codes. We have considered the sample of bright GRBs afterglows where the absorption continuum observed by XMM-PN shows strong indication of this kind of photoionization (Paper I). The XMM-PN is sensitive to the distinct shape of the absorption continuum but its energy resolution is not sufficient to detect the associated weak narrow lines.  We have then searched for the  lines predicted by the model in the RGS spectra. In the one case where the RGS would allow the detection of some of the predicted features GRB 090618 at z=0.54), we find a joint  $\approx$ 3 sigma evidence for complex narrow-line absorption, the most prominent of which imprinted by the ensemble of highly-ionized Fe L. While the statistical evidence is at $\approx 3 \sigma$, two independent results support the detection. First the parameters of the TEPID absorber derived by fitting the RGS data are consistent  with those derived from EPIC-PN. Then, if the redshift is left free to vary, the model recovers a value $z=0.540\pm0.003$ in agreement with the optical one and with an error that is consistent with that expected by the energy resolution of the instrument.
	
	Using the lines predicted by TEPID in the XMM sample as reference, we have first explored the possibility to detect ionized absorption lines with \textit{XRISM}/Resolve for a relatively bright afterglow. As discussed in Paper I, our sample is not biased toward luminous or unobscured GRBs, and thus the range of metal line intensities presented in Fig. \ref{fig:EW_XIFU_XRISM} can be considered representative of the broader GRB population. For the current (gate-valve closed) configuration we find that there are good prospect for a detection of the {\it joint} contribution of ionized Fe-K lines (rest frame energy$\approx 7 $ keV) for bright afterglows at $z\lesssim1$. We estimate that about 3 such bright afterglows per year can be observed by \textit{XRISM} assuming a TOO observation starting 1 day after the burst with an exposure of 100 ksec. The perspectives improve significantly, should the gate valve be successfully open. In such case, lines from other elements would become accessible, for a wider range of redshift. 
	
	With its exquisite sensitivity to weak lines, the X-IFU instrument \citep{2023ExA....55..373B, 2025ExA....59...18P} onboard \textit{NewAthena} promises to be a game-changer. We have extended the XMM sample including two representative cases of fainter GRB at higher redshifts, upto z=7, adopting as fiducial values of the parameter their mean values derived from the XMM sample. Fig.\ref{fig:EW_XIFU_XRISM} shows that most of the lines predicted by TEPID should be detectable by XIFU. We estimate that a dedicated follow up program by \textit{NewAthena}/XIFU can lead to a sample of upto 200 afterglows in 4 years with the required fluence, assuming the nominal TOO observation starting 12 hours after the burst and a 50 ksec exposure.
	
	To assess quantitatively the capability of the \textit{NewAthena}/XIFU to carry out a systematic measurement of the key parameters of the absorption medium in particular density, size, redshift,and metal abundances for a wide population of GRB,  we have carried out extensive simulations of the afterglow sample above as would be observed by the XIFU instrument. We find that  the density and the column density (or the size) of the star-forming region embedding the GRB can be measured with a error of $\approx 0.05$ dex from close to high redshift z=7 GRBs. Furthermore, the redshift can be immediately derived with a relative error $\lesssim 10^{-3}$ (Tab. \ref{tab:summary}). This is particularly important considering that while more than 90\% have an X-ray afterglow about 50\% are optically {\it dark} \citep{Piro2002,Jakobsson+04,Depasquale+2006}. Therefore, enabling X-ray redshift measurements would get rid of a critical selection bias on the population of GRB with measured redshift.
	
	The possibility of measuring metallicity by X-ray absorption in the 0.2-10 keV range with \textit{NewAthena} is effectively enabled by the possibility to constrain independently the metal column densities, through their narrow absorption features, and the presence of absorption by  H and He not fully ionized, that can be significant up to 1 keV rest-frame. We find that metallicity can be measured with a relative error ranging from $\lessapprox 10\%$ to $\approx 20\%$ from close to z$\approx 3$ bursts, while it is unconstrained for the burst at z=7  due to the marginal effect of H and He absorption shifted at the lower limit of the X-ray band.
	
	We went a step further to investigate the extent to which the abundance patterns of key metals can be constrained, given that different progenitors leave distinct chemical signatures. We find that the individual initial abundances of S, Si, and Fe are recovered within $\approx 10-30\%$ uncertainties.
	
	Finally we investigated the capability of X-IFU to distinguish among pristine, Pop~II, and Pop~III-enriched environments at high redshift. Pristine media exhibit only low-energy H and He absorption, whereas low-density Pop~II environments ($\log n_0 \lesssim 2~\mathrm{cm^{-3}}$) can mimic pristine gas only under unphysically low densities compared to JWST star-formation constraints ($\log n_0 \gg 3~\mathrm{cm^{-3}}$). Crucially, top-heavy Pop~III environments can be uniquely identified by the presence of Si and S absorption lines combined with a prominent absence of Fe features, yielding an upper limit of $\mathrm{Fe/(Si,S)} \lesssim 0.15$ compared to the unity ratio typical of Pop~II media.
	
	In conclusion we expect that \textit{NewAthena}/X-IFU will routinely provide high-resolution spectra for a vast number of GRB afterglows, mapping the properties of the star-forming medium embedding the GRB progenitor as a function of progenitor type and across cosmic time.

	%
	%
	
	\begin{acknowledgements}
		L.P., A.L.T., A.L. and F.N. acknowledge support from the European Union Horizon 2020 programme under the AHEAD2020 project (Grant Agreement No. 871158). A.L.T. and L.P. acknowledge support from the Italian Space Agency (Contract No. 2019-27-HH.0). A.L. and F.N. acknowledge support from the PRIN-MUR-2022 (Advanced X-ray modeling of black hole winds, DRAGON, ID 2022K9N5B4). Y.K. acknowledges support from Grant No. PAPIIT-UNAM IN102023. 
	\end{acknowledgements}
	
	\bibliographystyle{aa}
	\bibliography{references}
	
	\begin{appendix}
		
		\section{Additional tables and plots}\label{addplot}
		
		\begin{figure}
			\centering
			\includegraphics[width=\linewidth]{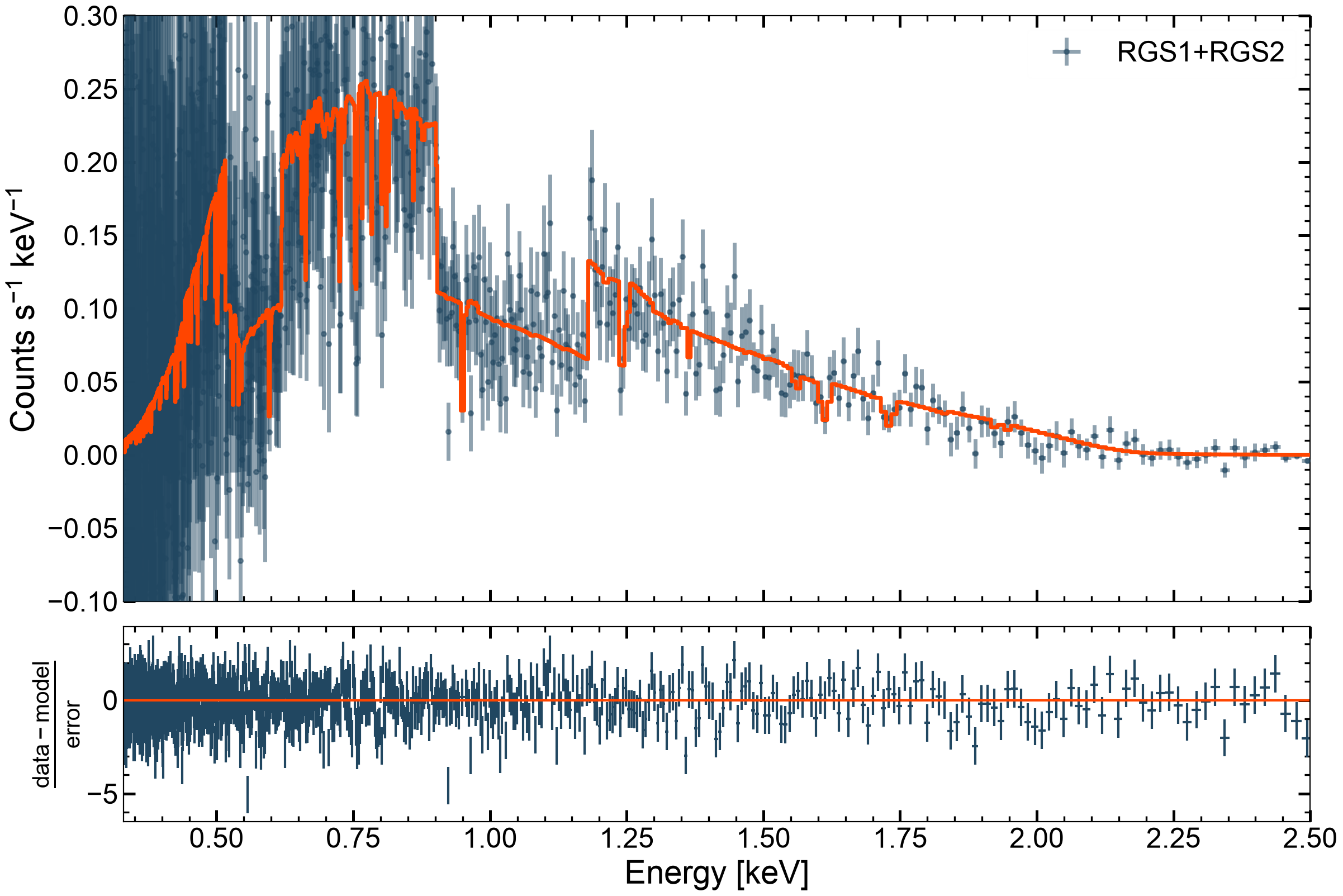}
			\caption{RGS1+RGS2 combined spectrum of GRB 090618 for the full fitted range. Overplotted in orange is the folded model for the TEPID best-fit. The bottom panel plots the residuals.}
			\label{fig:RGSfull090618}
		\end{figure}
		
		\begin{table*}[ht]
			\centering
			\begin{tabular}{cccc|cc|c|c}
				\hline\hline
				& \multicolumn{3}{c|}{RGS} & \multicolumn{2}{c|}{PN} & X-IFU & E$_{ionisation}$\\
				\hline
				GRB & Start time & Scheduled exposure & Effective exposure & PN effective & fluence & fluence &  \\ 
				& (T0+, ksec) & (ksec) & (ksec) & (ksec) & (erg cm$^{-2}$) & (erg cm$^{-2}$) & $10^{52}$ erg\\
				\hline
				060729 & 45 & 60.0 & 58.2 & 52.0 & 5.0$\times 10 ^{-7}$ & 4.9$\times 10 ^{-7}$ & 5.4  \\
				061121 & 22 & 40.0 & 38.3 & 30.0 & 1.7$\times 10 ^{-7}$ & 2.9$\times 10 ^{-7}$ & 19 \\
				080411 & 60 & 100.0 & 80.6 & 14.9 & 6.0$\times 10 ^{-8}$ & 2.0$\times 10 ^{-7}$ & 600 \\
				090618 & 19 & 23.0 & 18.3 & 8.60 & 1.3$\times 10 ^{-7}$ & 7.7$\times 10 ^{-7}$ & 70 \\
				120711A & 78 & 52.0 & 48.2 & 31.0 & 2.0$\times 10 ^{-7}$ & 3.3$\times 10 ^{-7}$ & 120 \\
				221009A[exp1] & 200 & 50.0 & 48.6 & - & - & - & - \\
				221009A[exp2] & 400 & 60.0 & 51.3 & 21.0 & 6.9$\times 10 ^{-7}$ & 1.2$\times 10 ^{-6}$ & 170 \\
				GRB at z=3 & - & - & - & - & - & 1.0$\times 10 ^{-7}$ & 18 \\
				GRB at z=7 & - & - & - & - & - & 1.0$\times 10 ^{-7}$ & 60 \\
				\hline
			\end{tabular}
			\caption{RGS observation log. Scheduled and effective exposures are for a single RGS first order spectrum. PN fluence is calculated as the average flux from the TEPID model multiplied by the effective PN exposure, X-IFU fluence assumes the same flux for an integration of 50 ksec. The ionisation energy is calculated up to the midpoint of the observation.}
			\label{tab:XMMlog}
		\end{table*}
		
		\begin{figure}[h]
			\centering
			\includegraphics[width=0.8\linewidth]{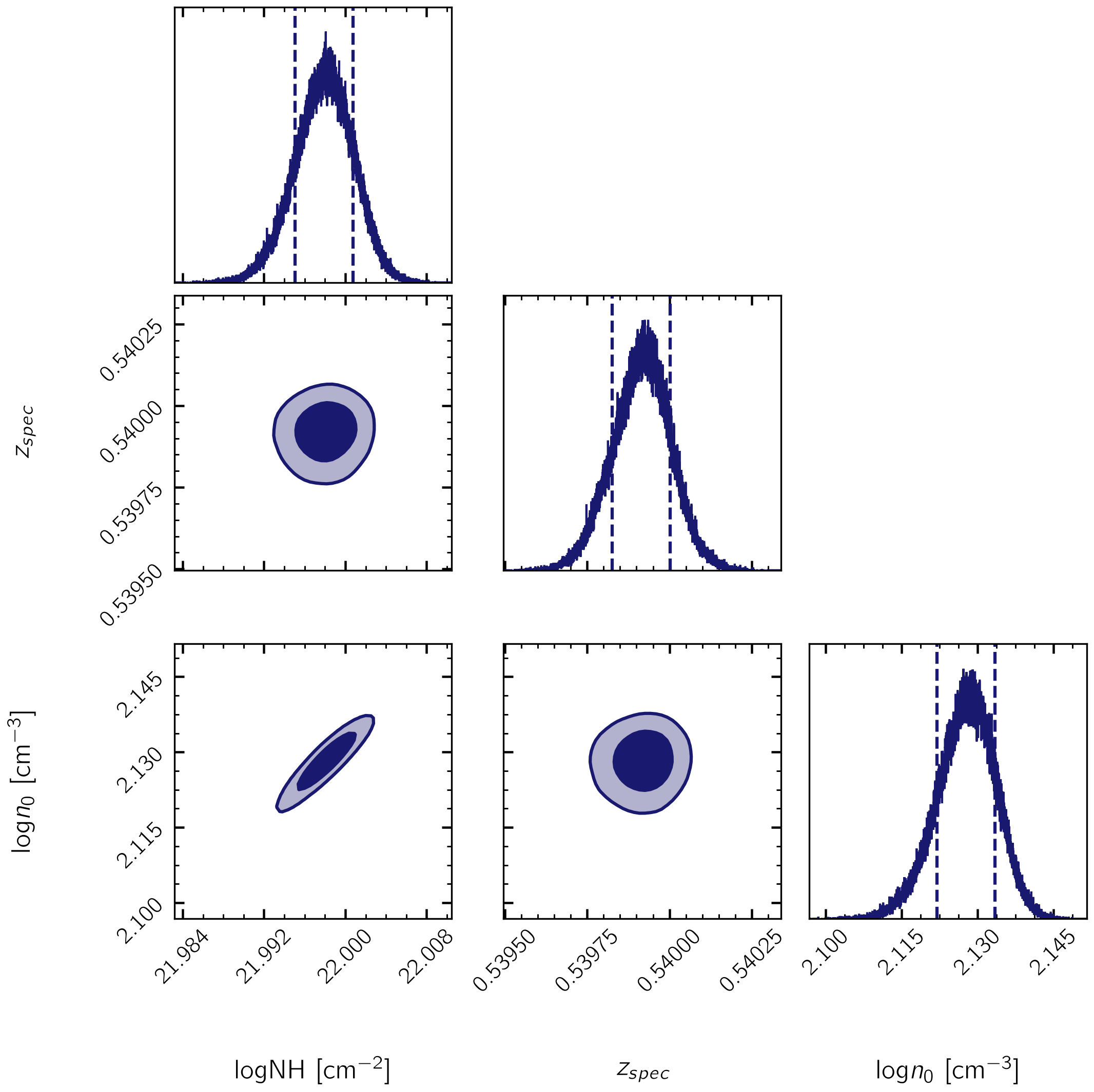}
			\caption{Contour plot obtained from the fit of the X-IFU spectrum of GRB 090618. The contours are plotted at the 1-$\sigma$ and 90\% confidence levels.}
			\label{fig:0906_fixed}
		\end{figure}
		
		\begin{figure}[h]
			\centering
			\includegraphics[width=0.8\linewidth]{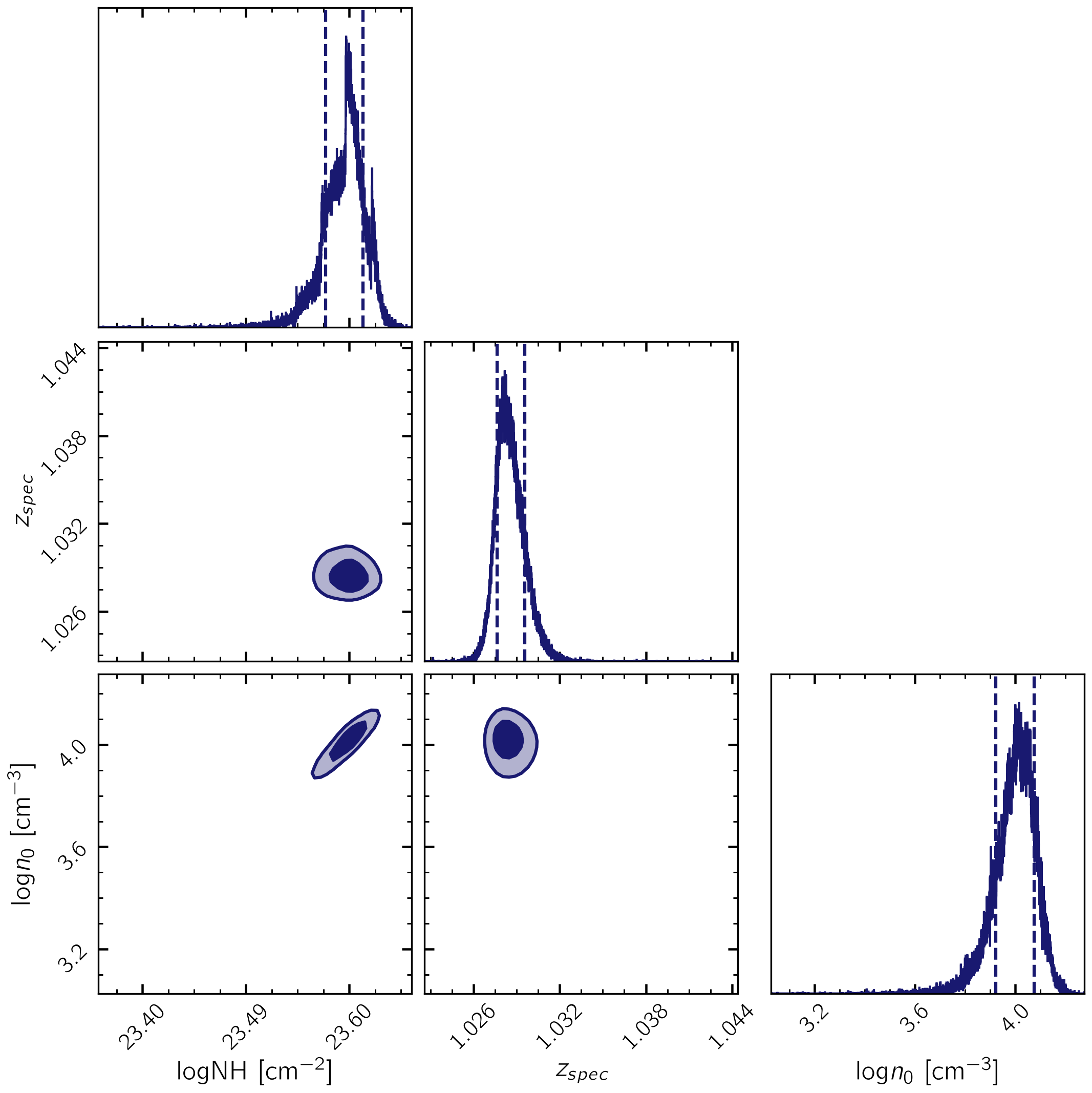}
			\caption{Contour plot obtained from the fit of the X-IFU spectrum of GRB 080411. The contours are plotted at the 1-$\sigma$ and 90\% confidence levels.}
			\label{fig:0804_fixed}
		\end{figure}
		
		\begin{figure}[h]
			\centering
			\includegraphics[width=0.8\linewidth]{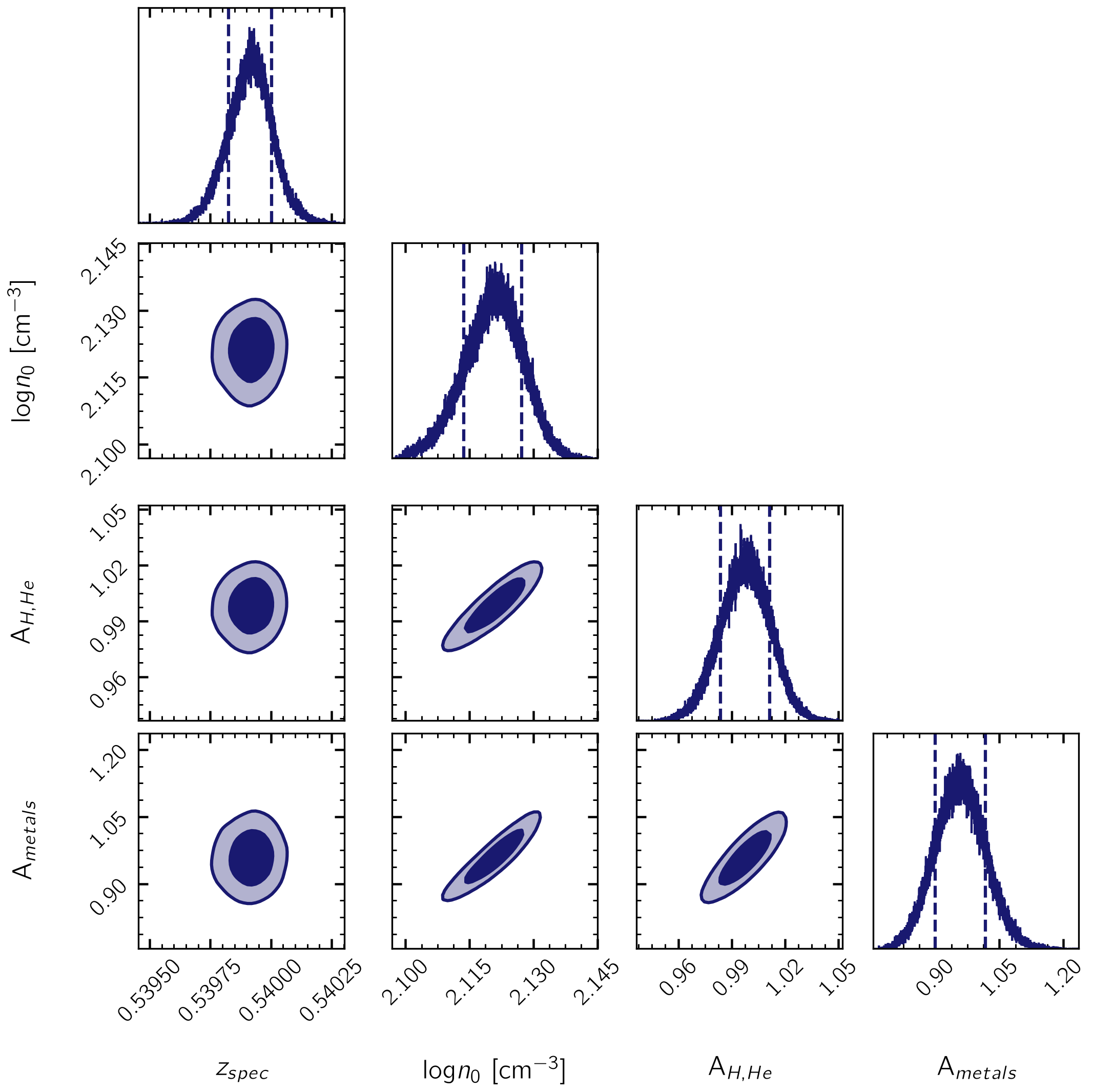}
			\caption{Contour plot obtained from the fit of the X-IFU spectrum of GRB 090618, by fixing the column and leaving the normalisations of H and He ($A_{H,He}$) and metal ($A_{metals}$) columns with respect to the input values free to vary. The contours are plotted at the 1-$\sigma$ and 90\% confidence levels.}
			\label{fig:0906_varmet}
		\end{figure}
		
		\begin{figure}[h]
			\centering
			\includegraphics[width=0.8\linewidth]{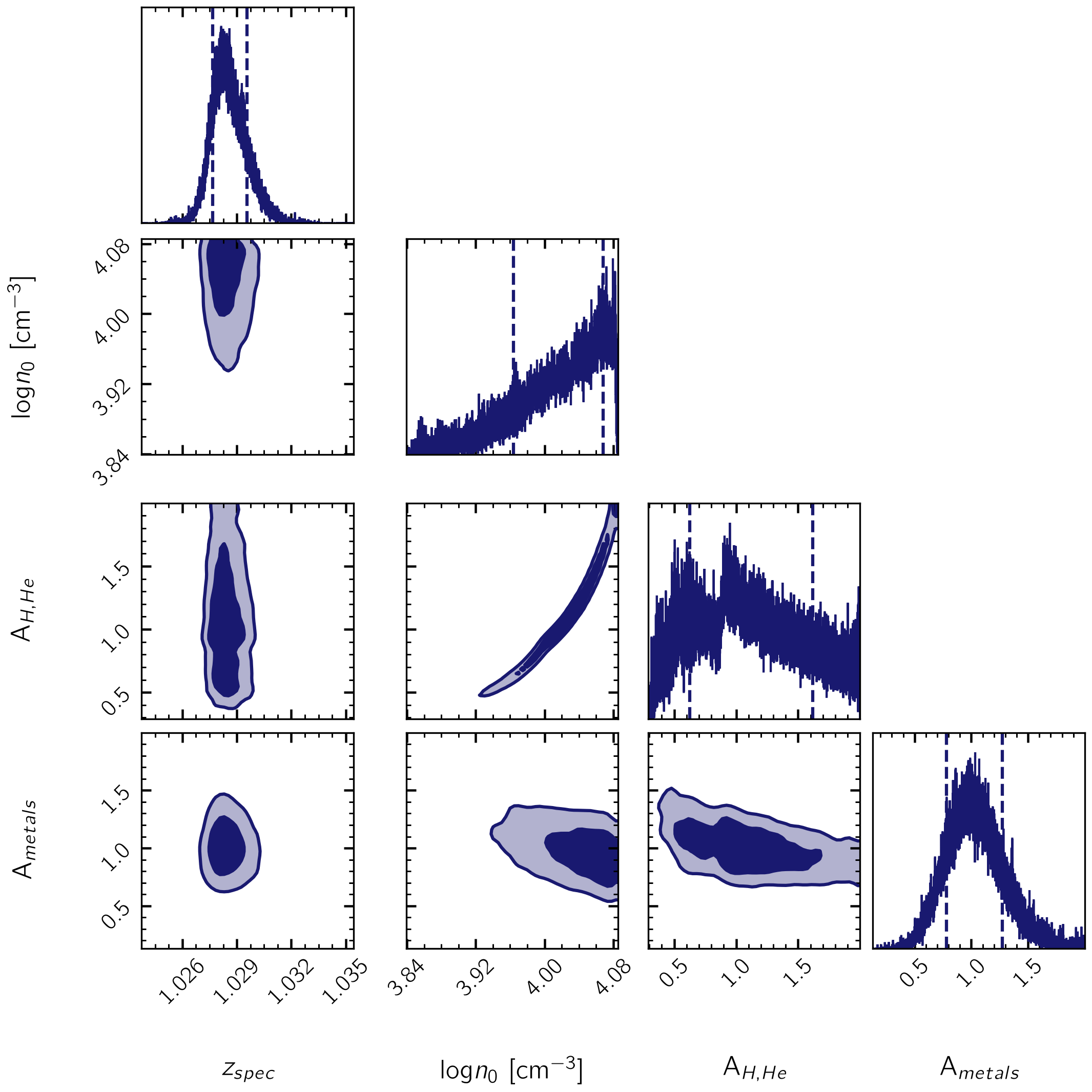}
			\caption{Contour plot obtained from the fit of the X-IFU spectrum of GRB 080411, by fixing the column and leaving the normalisations of H and He ($A_{H,He}$) and metal ($A_{metals}$) columns with respect to the input values free to vary. The contours are plotted at the 1-$\sigma$ and 90\% confidence levels.}
			\label{fig:0804_varmet}
		\end{figure}
		
		\begin{figure}[h]
			\centering
			\includegraphics[width=0.6\linewidth]{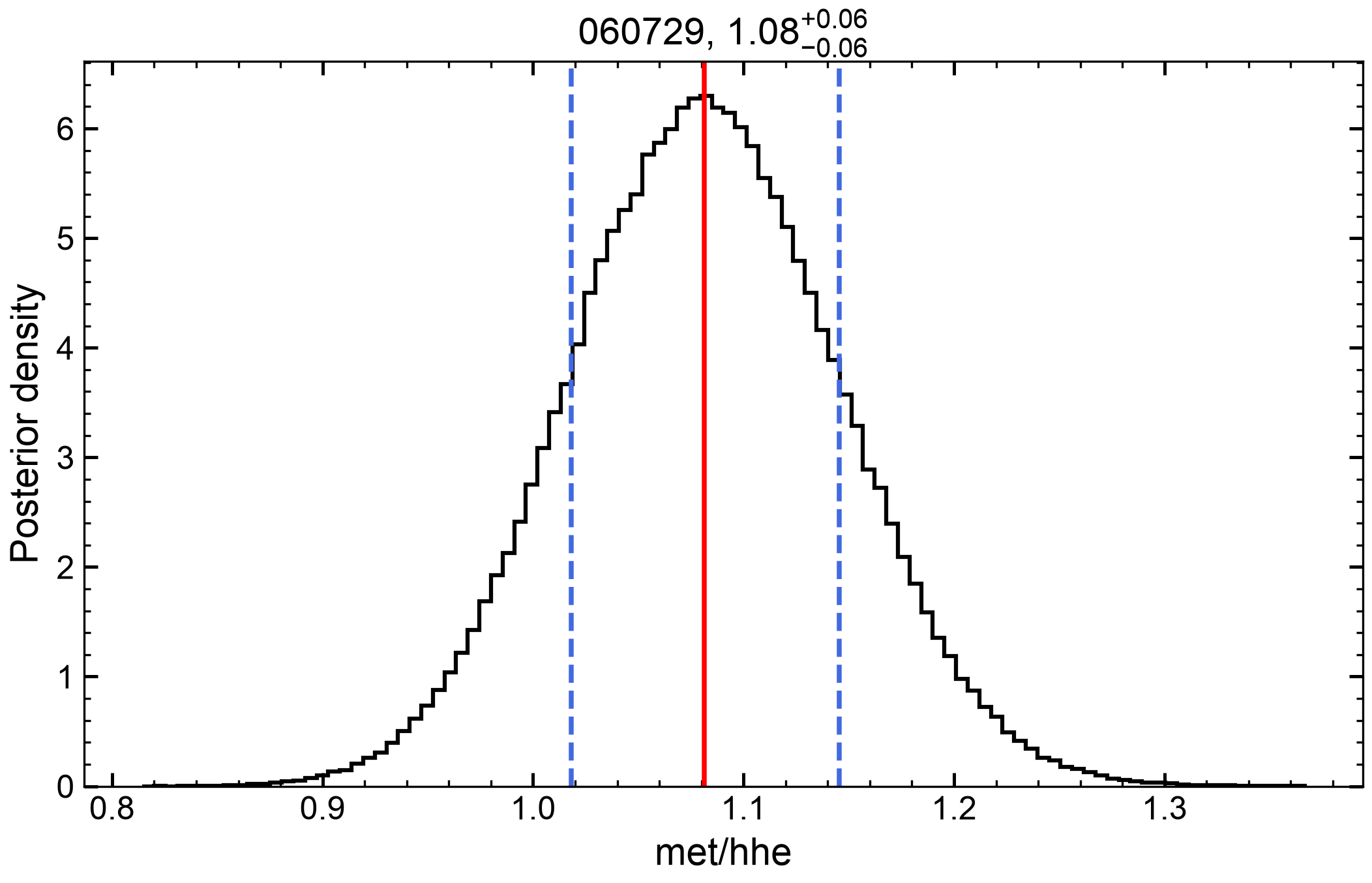}
			\caption{Posterior density of the metallicity obtained by fitting the X-IFU spectrum of GRB 060729. The metallicity is calculated as the ratio of $A_{metals}$ and $A_{H,He}$.}
			\label{fig:met060729}
		\end{figure}
		
		\begin{figure}[h]
			\centering
			\includegraphics[width=0.6\linewidth]{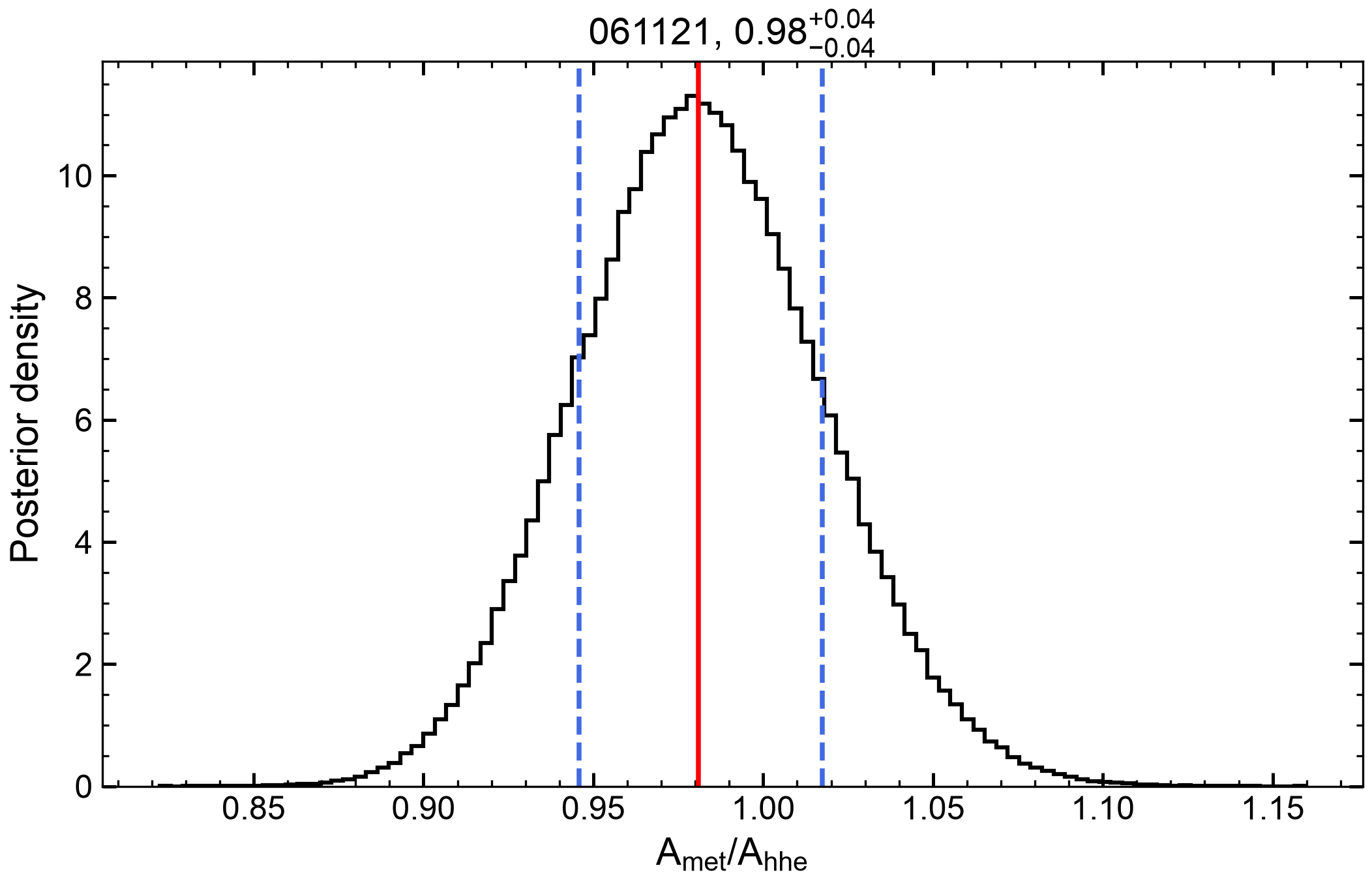}
			\caption{Posterior density of the metallicity obtained by fitting the X-IFU spectrum of GRB 061121. The metallicity is calculated as the ratio of $A_{metals}$ and $A_{H,He}$.}
			\label{fig:met061121}
		\end{figure}
		
		\begin{figure}[h]
			\centering
			\includegraphics[width=0.6\linewidth]{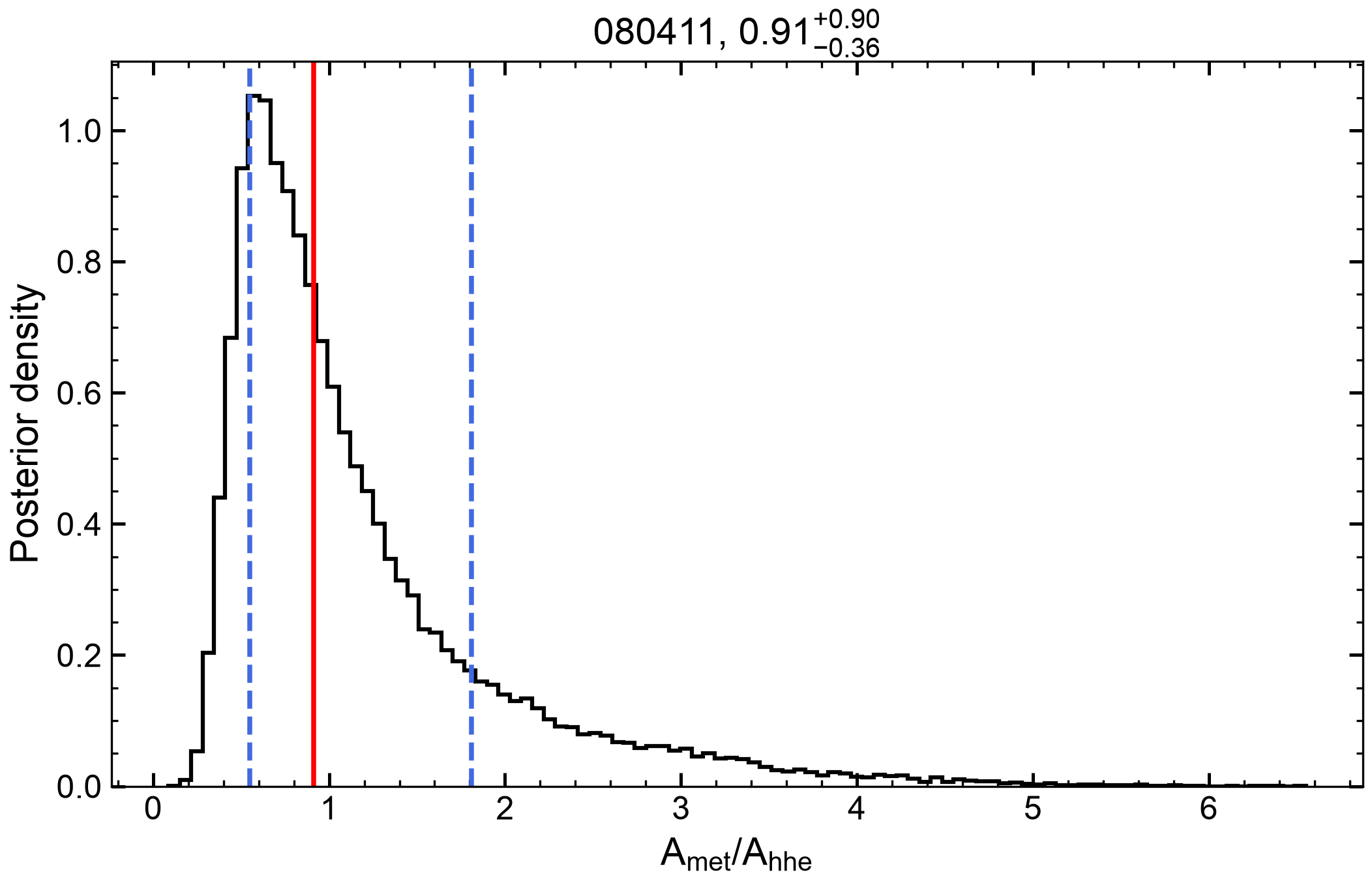}
			\caption{Posterior density of the metallicity obtained by fitting the X-IFU spectrum of GRB 080411. The metallicity is calculated as the ratio of $A_{metals}$ and $A_{H,He}$.}
			\label{fig:met080411}
		\end{figure}
		
		\begin{figure}[h]
			\centering
			\includegraphics[width=0.6\linewidth]{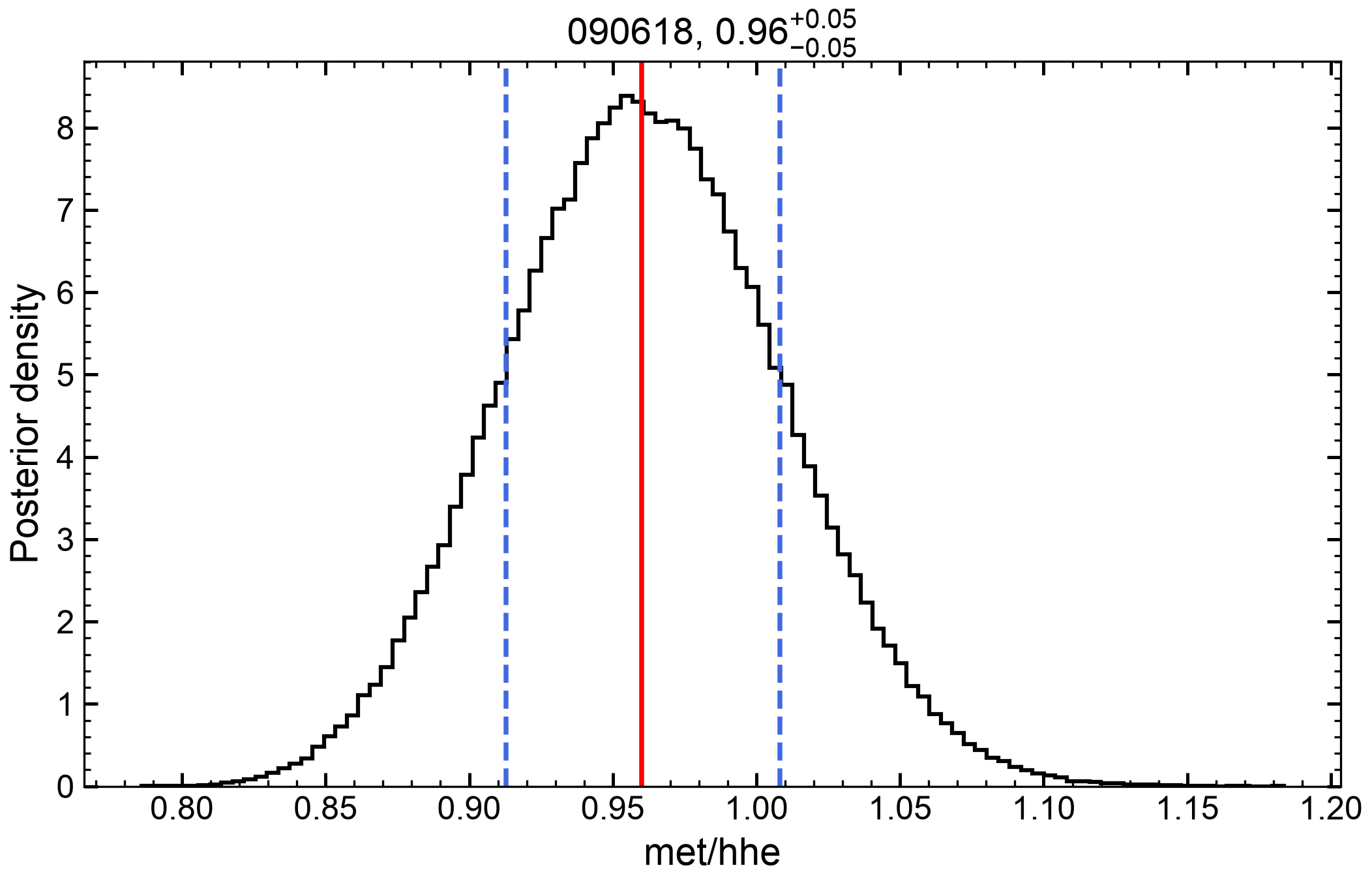}
			\caption{Posterior density of the metallicity obtained by fitting the X-IFU spectrum of GRB 090618. The metallicity is calculated as the ratio of $A_{metals}$ and $A_{H,He}$.}
			\label{fig:met090618}
		\end{figure}
		
		\begin{figure}[h]
			\centering
			\includegraphics[width=0.6\linewidth]{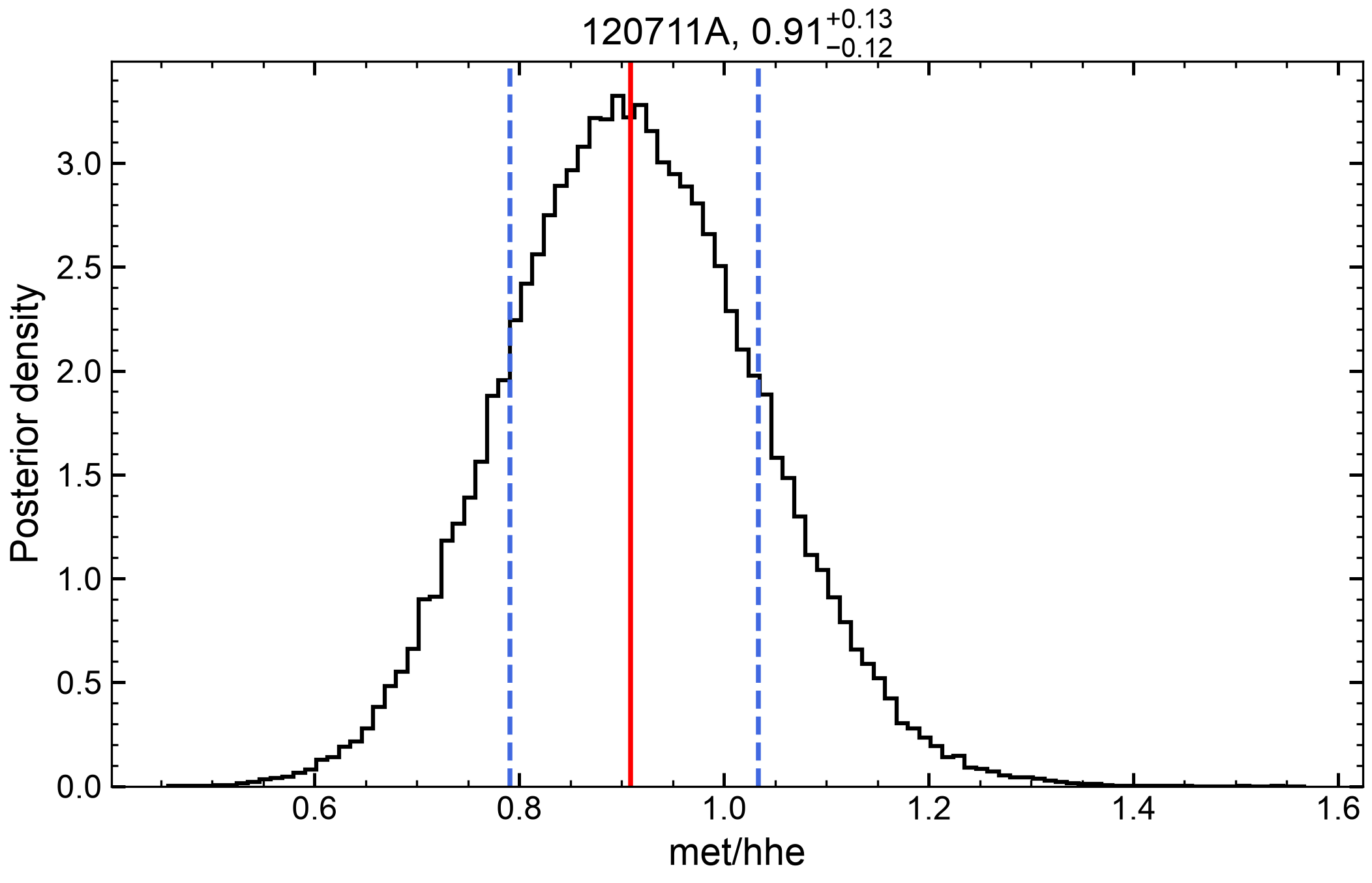}
			\caption{Posterior density of the metallicity obtained by fitting the X-IFU spectrum of GRB 120711A. The metallicity is calculated as the ratio of $A_{metals}$ and $A_{H,He}$.}
			\label{fig:met120711A}
		\end{figure}
		
		\begin{figure}[h]
			\centering
			\includegraphics[width=0.6\linewidth]{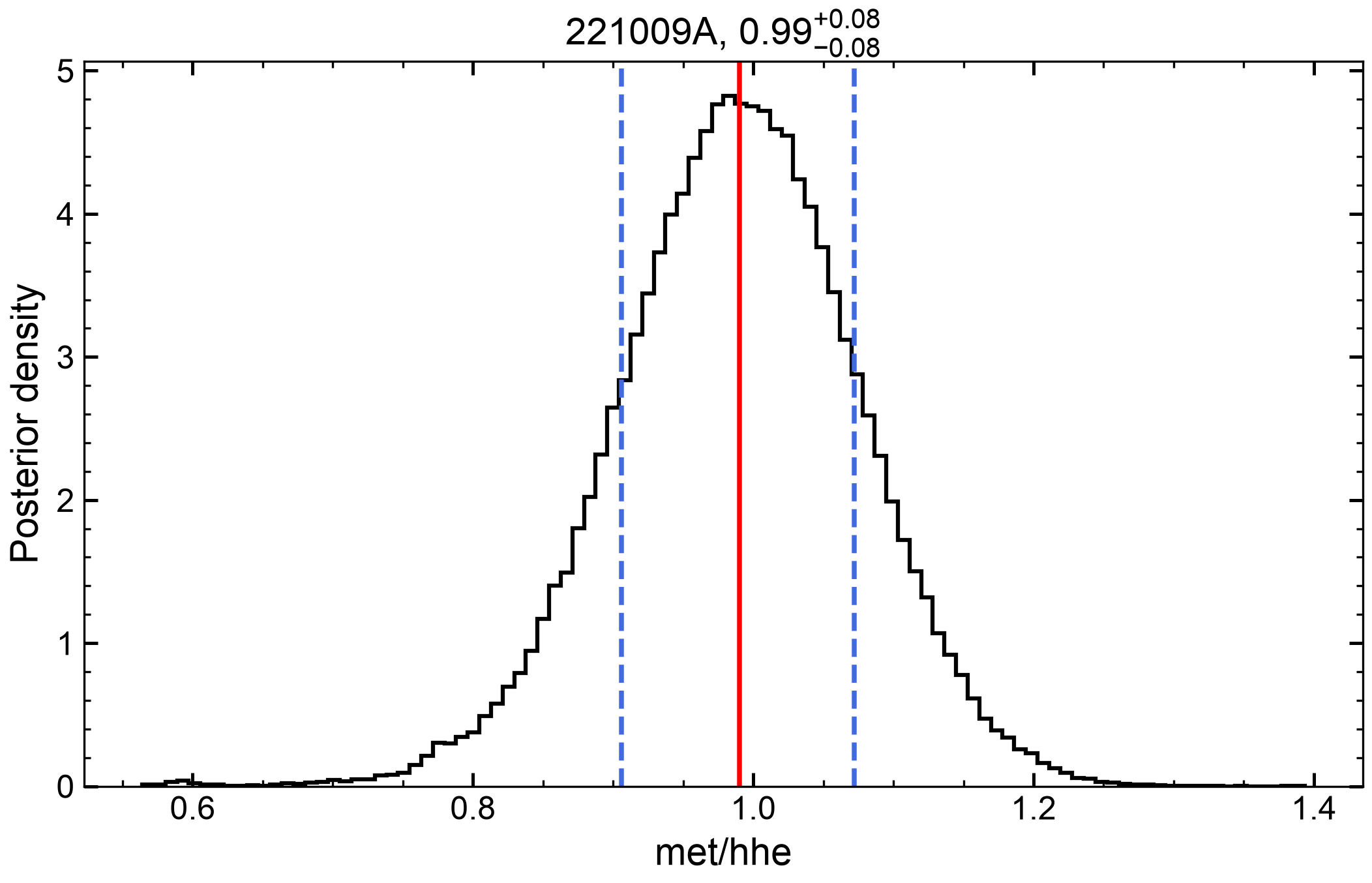}
			\caption{Posterior density of the metallicity obtained by fitting the X-IFU spectrum of GRB 221009A. The metallicity is calculated as the ratio of $A_{metals}$ and $A_{H,He}$.}
			\label{fig:met221009A}
		\end{figure}
		
		\begin{figure}[h]
			\centering
			\includegraphics[width=0.6\linewidth]{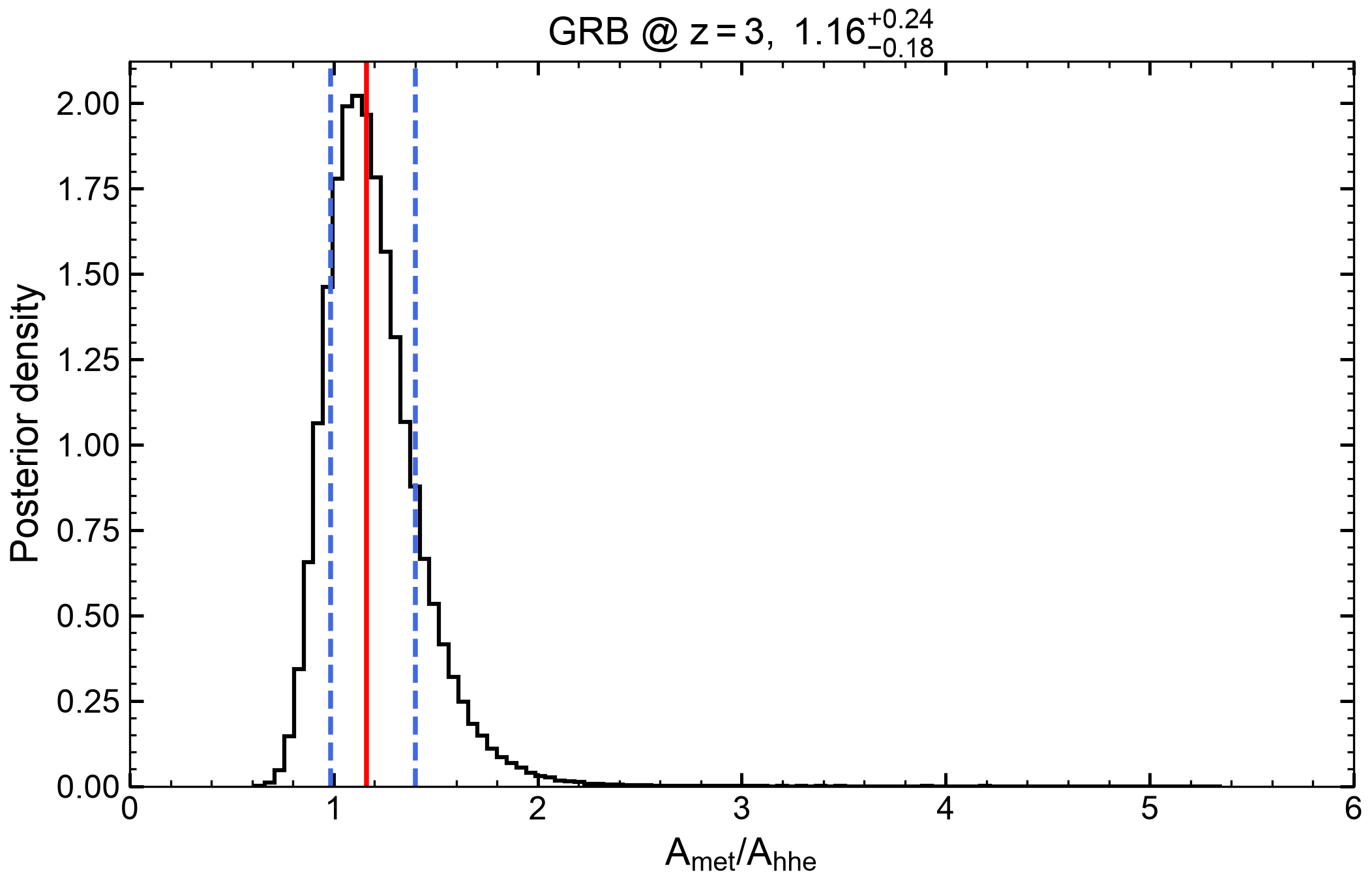}
			\caption{Posterior density of the metallicity obtained by fitting the X-IFU spectrum of a PopII GRB at z=3. The metallicity is calculated as the ratio of $A_{metals}$ and $A_{H,He}$.}
			\label{fig:metz3}
		\end{figure}
		
		\begin{figure}[h]
			\centering
			\includegraphics[width=0.8\linewidth]{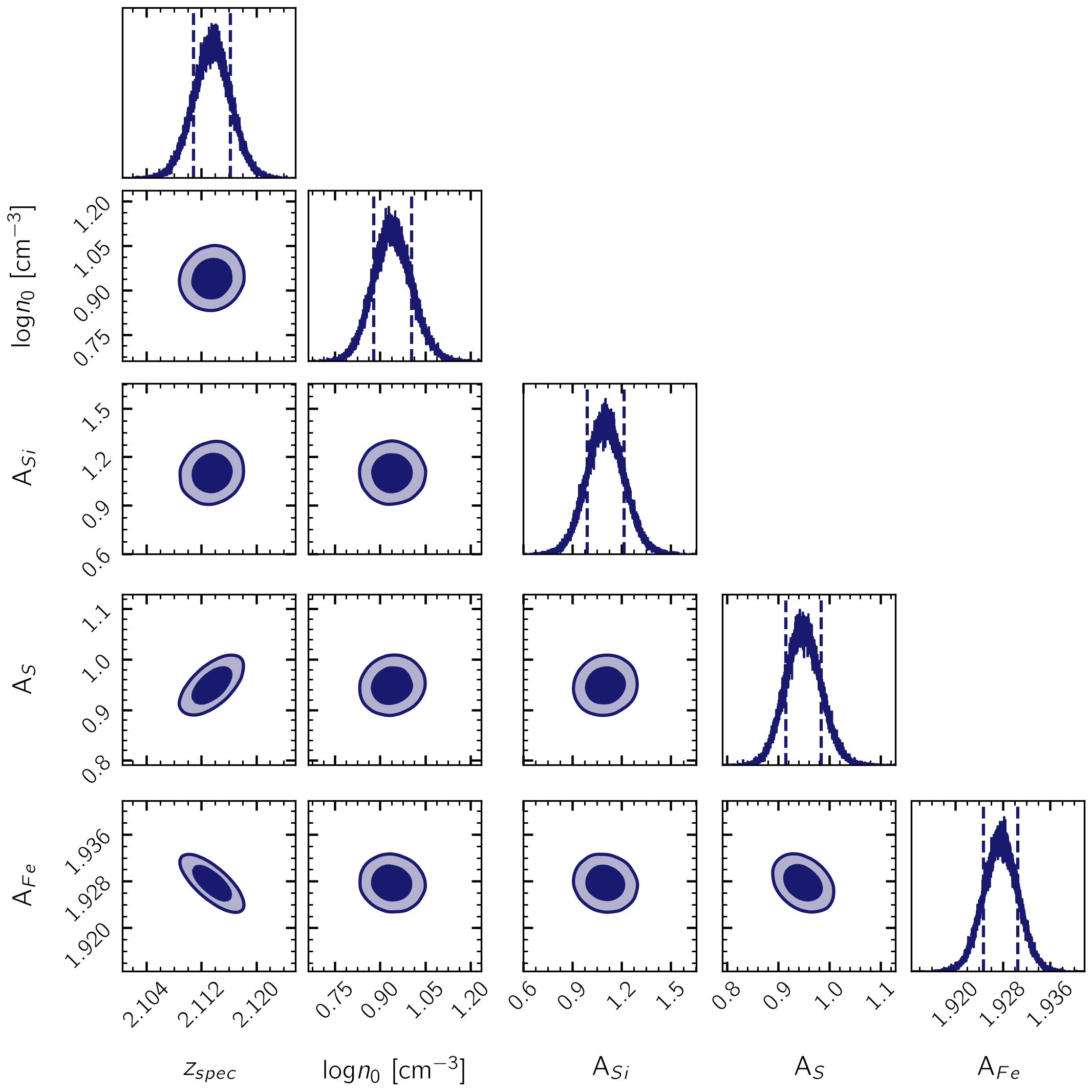}
			\caption{Contour plot obtained from the fit of the X-IFU spectrum of GRB 090618, by fixing the column and leaving the normalisations of Si ($A_{Si}$), S ($A_{S}$) and Fe ($A_{Fe}$) columns with respect to the input values free to vary. The contours are plotted at the 1-$\sigma$ and 90\% confidence levels.}
			\label{fig:0906_fracs}
		\end{figure}
		
		\begin{figure}[h]
			\centering
			\includegraphics[width=0.8\linewidth]{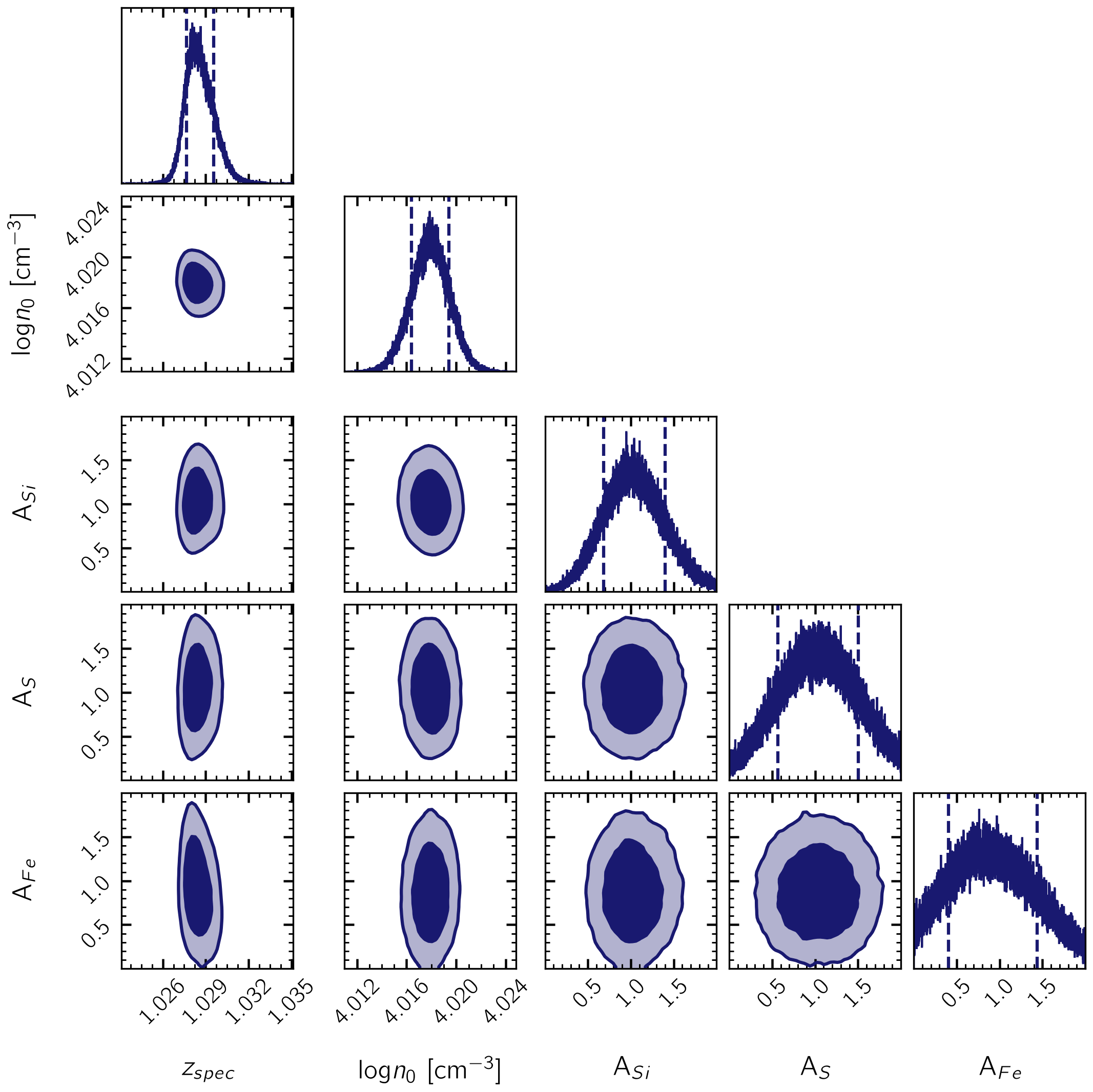}
			\caption{Contour plot obtained from the fit of the X-IFU spectrum of GRB 080411, by fixing the column and leaving the normalisations of Si ($A_{Si}$), S ($A_{S}$) and Fe ($A_{Fe}$) columns with respect to the input values free to vary. The contours are plotted at the 1-$\sigma$ and 90\% confidence levels.}
			\label{fig:0804_fracs}
		\end{figure}
		
		\begin{figure}[h]
			\centering
			\includegraphics[width=0.8\linewidth]{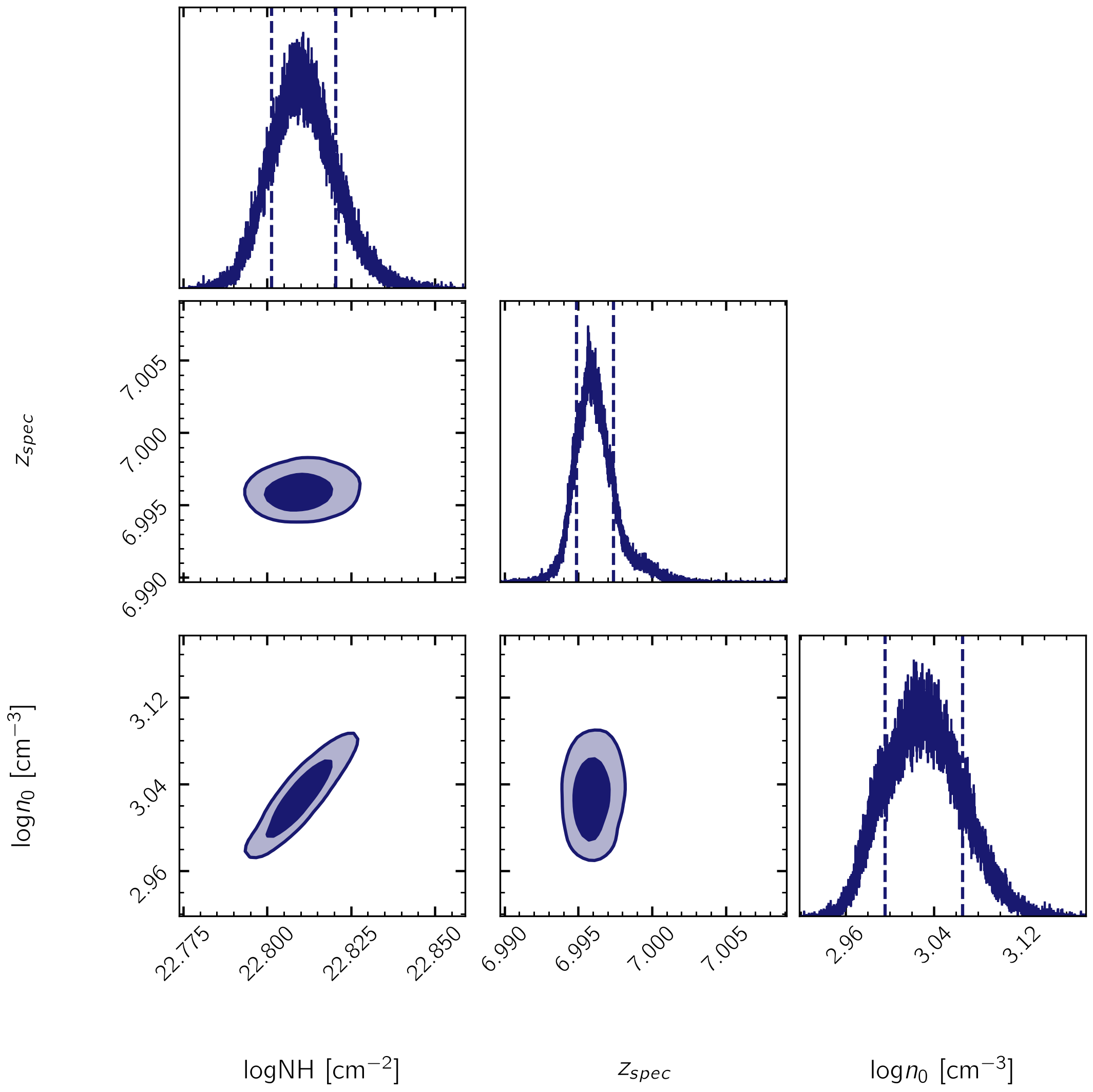}
			\caption{Contour plot obtained from the fit of the X-IFU spectrum of a PopII GRB at z=7. The contours are plotted at the 1-$\sigma$ and 90\% confidence levels.}
			\label{fig:z7_fixed}
		\end{figure}
		
		\begin{figure}[h]
			\centering
			\includegraphics[width=0.8\linewidth]{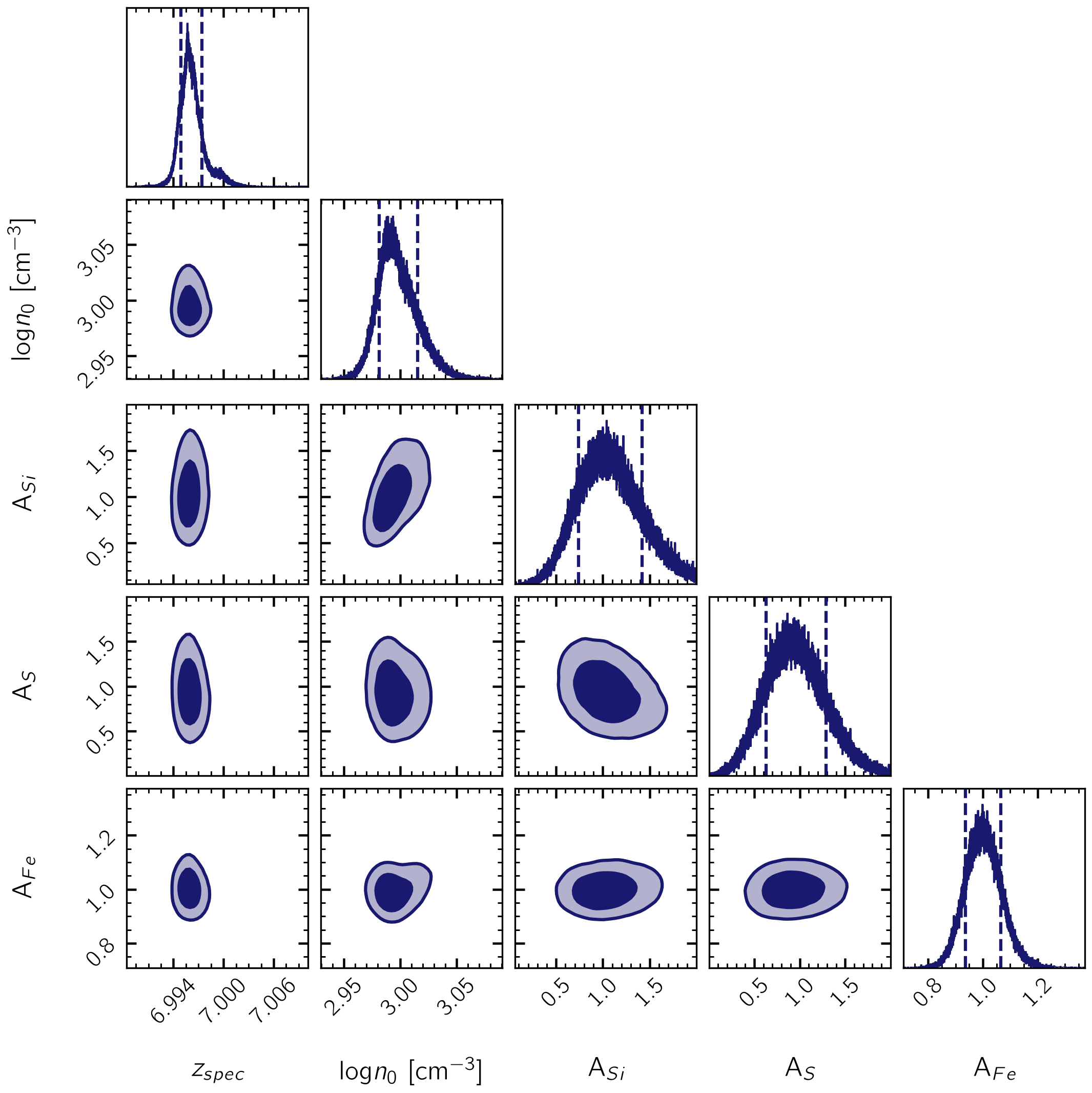}
			\caption{Contour plot obtained from the fit of the X-IFU spectrum of a PopII GRB at z=7 by fixing the column and leaving the normalisations of Si ($A_{Si}$), S ($A_{S}$) and Fe ($A_{Fe}$) columns with respect to the input values free to vary. The contours are plotted at the 1-$\sigma$ and 90\% confidence levels.}
			\label{fig:z7_fracs}
		\end{figure}
		
		\begin{figure}[h]
			\centering
			\includegraphics[width=0.8\linewidth]{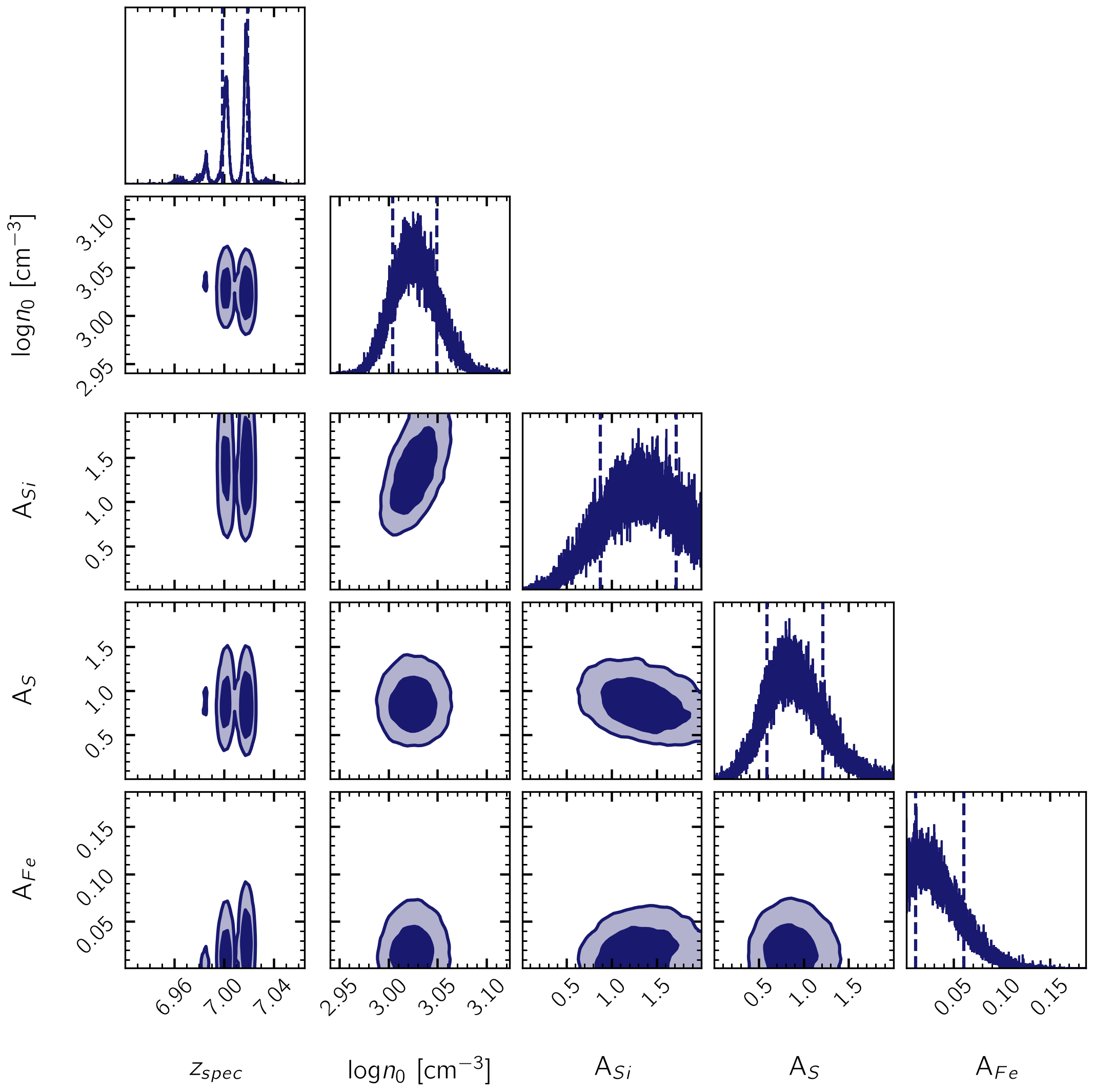}
			\caption{Contour plot obtained from the fit of the X-IFU spectrum of a PopIII-enriched PopII GRB at z=7 by fixing the column and leaving the normalisations of Si ($A_{Si}$), S ($A_{S}$) and Fe ($A_{Fe}$) columns with respect to the input values free to vary. The contours are plotted at the 1-$\sigma$ and 90\% confidence levels.}
			\label{fig:z7th_fracs}
		\end{figure}

		\begin{figure}[h]
			\centering
			\includegraphics[width=0.8\linewidth]{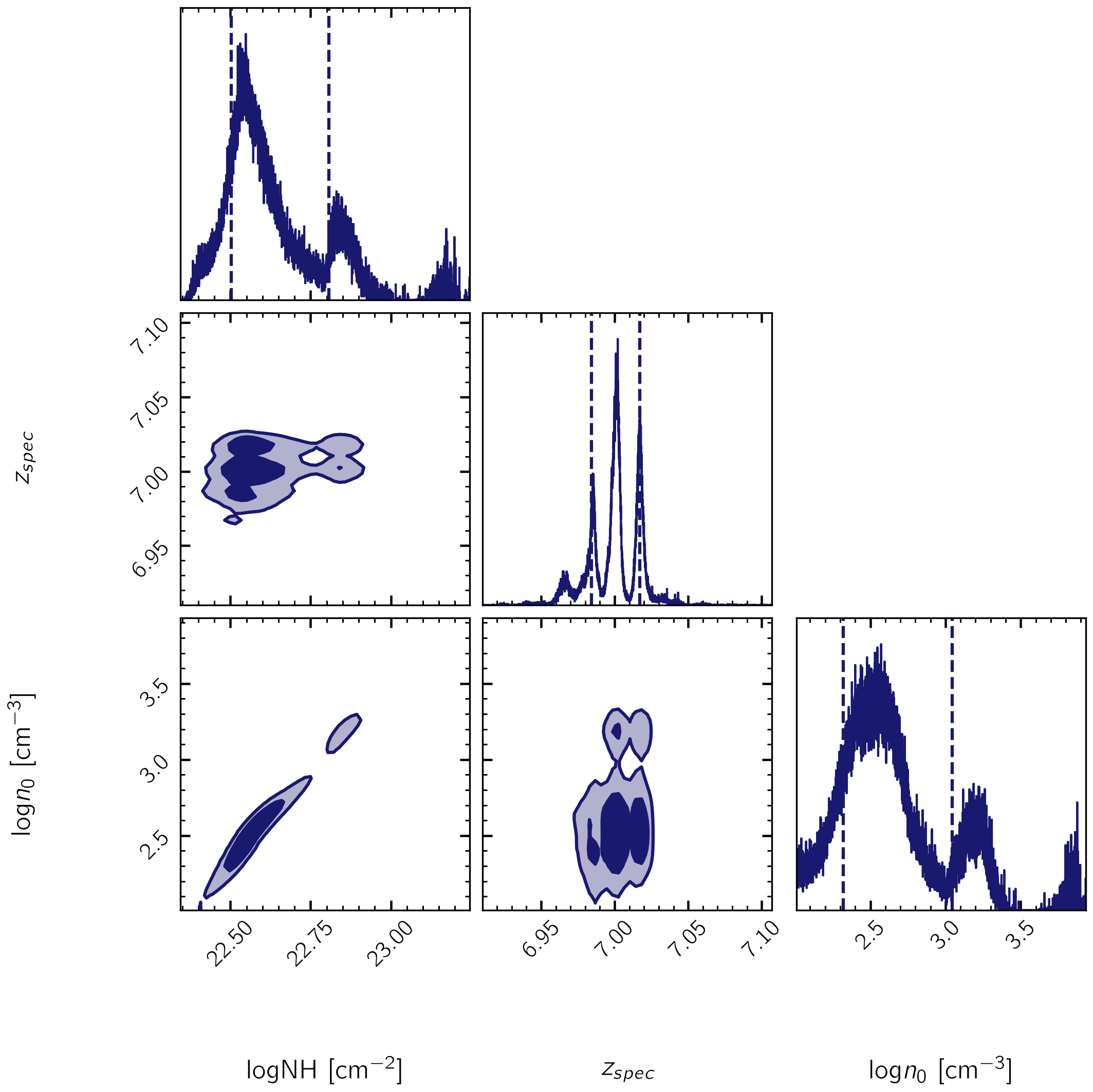}
			\caption{Contour plot obtained from the fit of the X-IFU spectrum of PopIII-enriched PopII GRB at z=7. The contours are plotted at the 1-$\sigma$ and 90\% confidence levels.}
			\label{fig:z7th_fixed}
		\end{figure}
		
	\end{appendix}
	
\end{document}